\documentclass[11pt]{article}
\pdfoutput=1
\usepackage{jcappub,natbib}
\usepackage[normalem]{ulem}
\definecolor{verde}{rgb}{0,0.5,0}

\graphicspath{{./figures/}}

\def\la{\mathrel{\raise.3ex\hbox{$<$\kern-.75em\lower1ex\hbox{$\sim$}}}}
\usepackage{orcidlink}

\hypersetup{
colorlinks   = true, 
urlcolor     = blue, 
linkcolor    = blue, 
citecolor   = blue 
}
\usepackage[capitalise]{cleveref}
\usepackage[normalem]{ulem} 
\begin{document}
\title{Kinetic Fragmentation of the QCD Axion on the Lattice}

\author[a]{Matteo Fasiello\orcidlink{0000-0002-2532-5202},}
\author[b,c]{Joanes Lizarraga\orcidlink{0000-0002-1198-3191},}
\author[a,1]{Alexandros Papageorgiou\orcidlink{0000-0002-2736-3026}, \note{corresponding author}}
\author[b,c]{and Ander Urio\orcidlink{0000-0002-0238-8390}}

\affiliation[a]{Instituto de F\'{i}sica T\'{e}orica UAM-CSIC, c/ Nicol\'{a}s Cabrera 13-15, 28049, Madrid, Spain}
\affiliation[b]{Department of Physics, University of the Basque Country, UPV/EHU, 48080, Bilbao, Spain}
\affiliation[c]{EHU Quantum Center, University of the Basque Country UPV/EHU, Leioa, 48940 Biscay, Spain}

\emailAdd{matteo.fasiello@csic.es}
\emailAdd{joanes.lizarraga@ehu.eus}
\emailAdd{papageorgiou.hep@gmail.com}
\emailAdd{ander.urio@ehu.eus}
\abstract{Kinetic misalignment, one of the most compelling scenarios for the non-thermal generation of axion dark matter, is generally accompanied by axion fragmentation, a process in which the energy of the axion condensate is transferred to its perturbations. The dynamics of fragmentation, at least in the context of dark matter production, have so far been studied semi-analytically using perturbation theory. In this work, we present the first classical lattice simulation of kinetic axion fragmentation in the context of dark matter production, focusing on parameters relevant to the QCD axion.~Our findings indicate that the non-perturbative dynamics captured by the lattice lead to a significantly broader spectrum of axion fluctuations, with a sustained transfer of energy to mildly relativistic modes and with smaller occupation numbers compared to the linear approximation.~As a consequence, the final dark matter abundance is typically ${\cal O}(1)$ lower than in the linear approximation, which is itself ${\cal O}(1)$ lower than the zero-mode-only prediction. This broadening and suppression of the spectrum could have a significant impact on axion mini-halo formation, one of the main experimental handles on kinetic fragmentation.}
\date{\today}
\maketitle
%
\section{Introduction}
\label{sec:intro}
%
%
The QCD axion, a pseudo-Nambu-Goldstone boson that arises as a consequence of a spontaneously broken chiral global $U(1)$ symmetry, is widely considered a prime candidate for resolving the strong CP problem of QCD \cite{Peccei:1977hh,Peccei:1977ur,Weinberg:1977ma,Wilczek:1977pj}. In addition to that, Axion-Like-Particles (ALPs) appear in many extensions of the standard model and are quite ubiquitous in string theory \cite{Gelmini:1980re,Davidson:1981zd,Wilczek:1982rv,Dienes:1999gw,Arvanitaki:2009fg,Cicoli:2013ana}. Moreover, axions are attractive dark matter candidates which can be produced by a variety of mechanisms. The two most commonly studied mechanisms for axion dark matter generation is i) non-thermal generation through the vacuum misalignment mechanism \cite{Preskill:1982cy,Abbott:1982af,Dine:1982ah,Higaki:2015jag,Higaki:2016yqk} and ii) generation through the decay of topological defects \cite{Kibble:1976sj,Kibble:1980mv,Davis:1985pt,Davis:1986xc,Harari:1987ht,Battye:1993jv,Higaki:2016jjh}.

The dynamics of the vacuum misalignment mechanism are straightforward to grasp. Assuming the Peccei-Quinn symmetry, which gives rise to the axion, to be broken before or sufficiently early during inflation, one expects the misalignment angle of the axion to take a random homogeneous value after inflation. Since at early times, the Hubble friction is large while the mass of the axion is small (due to the high temperature suppression of the instanton effects that give rise to the axion mass), the misalignment angle remains constant and the axion is frozen at some random point along its potential. At lower temperatures, when the mass of the axion and the Hubble scale are of the same order, the axion starts rolling down the potential eventually settling to a late time oscillation close to the potential minimum. The energy of the axion condensate scales in a matter-like way as $\rho_{\theta,0}\propto a^{-3}$ at late times and the final dark matter abundance depends on the axion mass $m_0$ and axion decay constant $f$ with a mild dependence on the initial misalignment angle $\theta_i$ as long as the angle is not fine-tuned close to the maximum or minimum of the axion potential.

\begin{figure}
    \centering
    \includegraphics[width=0.99\linewidth]{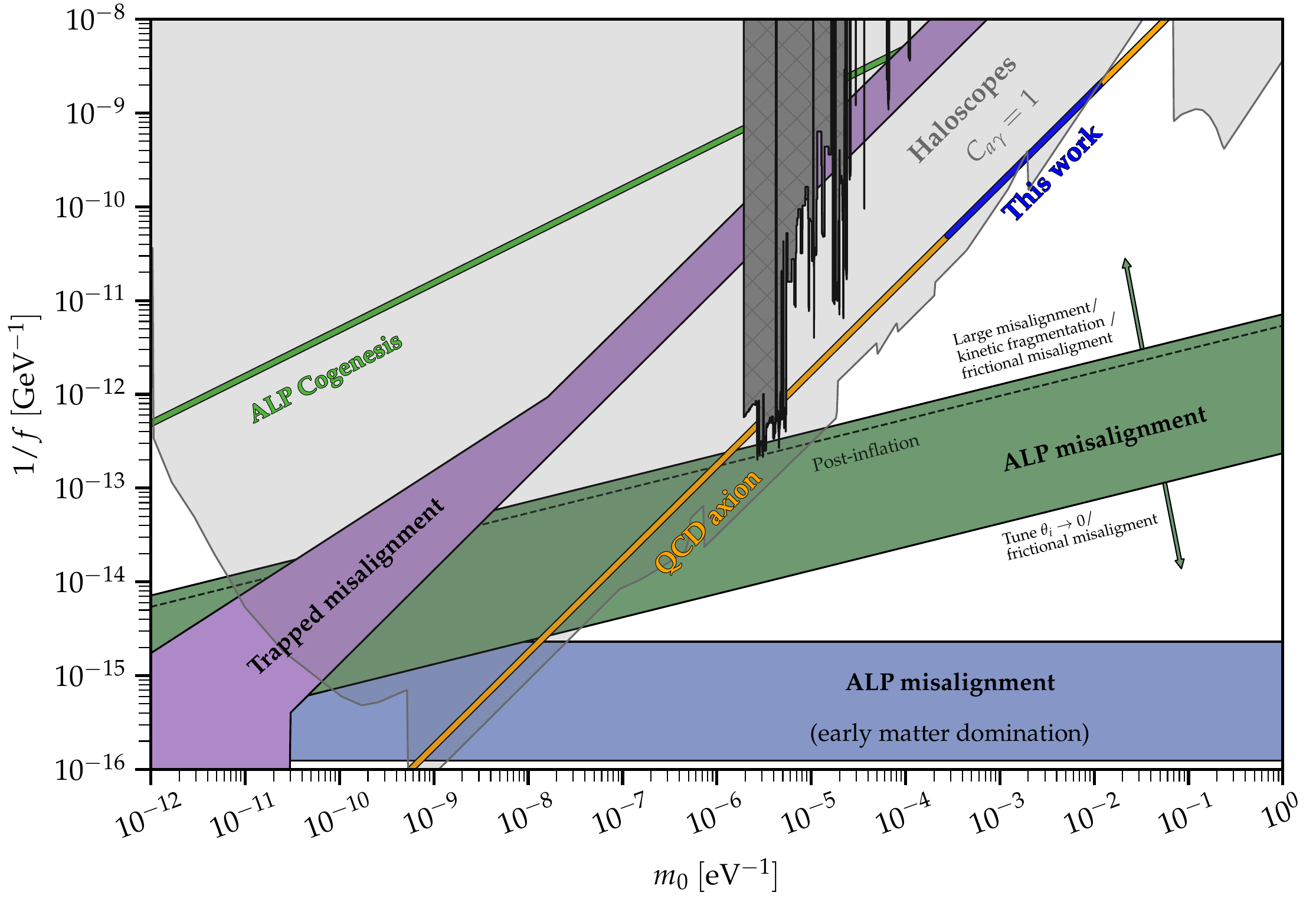}
    \caption{The landscape of the relevant parameter space for axion dark matter. The part of the QCD axion highlighted in blue is the parameter space studied in this work. The dark shaded region is the parameter space already excluded by haloscope experiments while the light gray area corresponds to the parameter space accessible by future haloscope experiments. The data in this plot has been adapted by \cite{AxionLimits}.}
    \label{fig:experiments}
\end{figure}

The vacuum misalignment mechanism can successfully account for the observed dark matter in a rather limited range of masses and axion decay constants which are rather challenging to probe experimentally. Additionally, in the case of the QCD axion, there is a very narrow window at around $f\simeq 10^{12}\;{\rm GeV}$ which can account for the observable dark matter abundance. Most of the ALP parameter space which is accessible by experiments such as haloscopes \cite{ADMX:2009iij,Caldwell:2016dcw,ADMX:2018gho,MADMAX:2019pub,ADMX:2019uok,CAPP:2020utb,ADMX:2021nhd,CAPP:2024dtx,BREAD:2021tpx} (see Fig.~\ref{fig:experiments}), generally lie in the regime where the axion decay constant is smaller than $f\simeq 10^{12}\;{\rm GeV}$. Such a small axion decay constant, at least in the case of the QCD axion would lead to under-production of axion dark matter if the means of its production is the vacuum misalignment mechanism. For this reason, there has been a tremendous effort in the literature to scrutinize and potentially modify the canonical vacuum misalignment mechanism.

Many alternative scenarios to the vacuum misalignment mechanism have been conceived such as a large \cite{Co:2018mho,Takahashi:2019pqf,Arvanitaki:2019rax,Huang:2020etx,Co:2024bme} or small \cite{Dvali:1995ce,Banks:1996ea,Choi:1996fs,Co:2018phi} initial misalignment angle, parametric resonance \cite{Co:2017mop,Harigaya:2019qnl,Co:2020dya}, frictional misalignment \cite{Papageorgiou:2022prc}, trapped misalignment \cite{Jeong:2022kdr,DiLuzio:2021gos,DiLuzio:2024fyt}, acoustic misalignment \cite{Bodas:2025eca}, bubble misalignment \cite{Lee:2024oaz}, varying axion decay constant \cite{Allali:2022yvx}, curvature induced dark matter \cite{Eroncel:2025qlk} interaction with monopoles \cite{Fischler:1983sc,Nakagawa:2020zjr}, level crossings \cite{Daido:2015cba,Murai:2024nsp}, modifications in the cosmological history of the universe \cite{Visinelli:2009kt,Nelson:2018via,Arias:2021rer,Dine:1982ah,Steinhardt:1983ia,Choi:2022btl,Dimopoulos:1988pw,Davoudiasl:2015vba,Hoof:2017ibo,Graham:2018jyp,Takahashi:2018tdu,Kitajima:2019ibn} and finally kinetic misalignment \cite{Co:2019jts,Chang:2019tvx,Co:2019wyp,Domcke:2020kcp,Co:2020jtv,Harigaya:2021txz,Chakraborty:2021fkp,Kawamura:2021xpu,Co:2021qgl,Co:2021lkc,Gouttenoire:2021wzu,Gouttenoire:2021jhk,Fonseca:2019ypl,Madge:2021abk,Eroncel:2022vjg,Eroncel:2022efc}. The large variety of plausible mechanisms for axion dark matter generation listed above lead to different predictions for the present day axion abundance compared to the standard vacuum misalignment scenario. As a result, it is important to probe experimentally the widest possible parameter space in the $(m_0,f)$ plane and not just the one associated with the QCD axion or the vacuum misalignment mechanism alone.

Among the aforementioned possibilities, the kinetic misalignment mechanism \cite{Co:2019jts} has attracted particular attention since it not only allows for the generation of axion dark matter, but it can also play a role in Baryogenesis \cite{Co:2019wyp,Domcke:2020kcp,Co:2020jtv,Co:2020xlh} and it generally points to smaller axion decay constants which are precisely the ones within reach of haloscope experiments in the near future. At the superficial level, kinetic misaligment is a modification of the initial conditions of the axion zero-mode at early times. Instead of the axion being initially frozen somewhere along the potential, one assumes that the axion has a large kinetic energy so that the potential is negligible by comparison. That allows for the axion to roll over many potential maxima, with its energy redshifting as $\rho_\theta\propto a^{-6}$, before it loses a sufficient amount of kinetic energy to be trapped by a potential minimum. If the trapping of the axion in a potential minimum takes place after the axion would have started to roll in the conventional vacuum misalignment, one ends up with a greater final axion dark matter abundance since the onset of oscillations has been delayed. Going beyond the superficial level, kinetic misalignment has been implemented in many UV completions which lead to a naturally large initial kinetic energy \cite{Co:2019jts,Eroncel:2024rpe,Lee:2024bij}.

Soon after the introduction of kinetic misalignment, it was pointed out that it is typically accompanied by axion fragmentation \cite{Fonseca:2019ypl,Chatrchyan:2020pzh,Morgante:2021bks}. The combined scenario that includes the two aforementioned effects is sometimes referred to as ``kinetic fragmentation" \cite{Eroncel:2022vjg,Eroncel:2022efc}. This occurs because as the axion zero mode rolls over potential maxima, or oscillates around a minimum with large amplitude, the axion perturbations experience tachyonic instabilities and/or parameteric resonance which exponentially enhances their occupation number. These enhanced perturbations then backreact on the axion, stripping away its energy. Fragmentation occurs when the energy in perturbations is much greater than the energy of the zero mode. Reference \cite{Eroncel:2022vjg} is the most comprehensive reference on the effects of fragmentation. The effects of fragmentation, at least for the purpose of axion dark matter generation, have so far been studied analytically or numerically using linear perturbation theory and homogeneous backreaction. The picture that emerges from this analysis is that fragmentation alters the final dark matter abundance compared to the standard kinetic misalignment by an ${\cal O}(1)$ factor. Additionally, due to the highly peaked production of axion perturbations, these perturbations have significant consequences for dark matter mini-halo formation at late times \cite{Xiao:2021nkb,Eroncel:2022efc}. This is particularly promising as it provides an experimental handle for falsifying this mechanism. Finally, the process of fragmentation can lead to production of gravitational waves whose amplitude can be greater than the gravitational wave background produced during inflation \cite{Eroncel:2022vjg}. While the scenario above is very appealing, it remains unclear how including non-perturbative effects and inhomogeneous backreaction may alter these predictions.

Studies of the axion-like particle (ALP) dark matter paradigm using non-perturbative lattice techniques have traditionally focused on the post-inflationary PQ symmetry breaking scenario, in which axions are predominantly produced via topological defects such as strings and domain walls (see \cite{Gorghetto:2018myk,Kawasaki:2018bzv,Vaquero:2018tib,Buschmann:2019icd,Klaer:2019fxc,Hindmarsh:2019csc,Gorghetto:2020qws,Gorghetto:2021fsn,Hindmarsh:2021vih,Buschmann:2021sdq,Baeza-Ballesteros:2023say,Benabou:2023ghl,Saikawa:2024bta,Kim:2024wku,Correia:2024cpk} for a collection of recent works). However, the application of such lattice methods to the study of axion fragmentation in the pre-inflationary scenario is an emerging area of interest, motivated by a variety of recent proposals. These include studies of the relaxion mechanism \cite{Morgante:2021bks}, the effect of monodromy \cite{Berges:2019dgr,Chatrchyan:2020pzh} or non-periodic potentials \cite{Chatrchyan:2023cmz}, and models involving couplings to hidden dark gauge sectors \cite{Ratzinger:2020oct,Weiner:2020sxn} and models involving multiple dark sectors \cite{Co:2025jnj}.

The aim of the present work is to perform the first comprehensive fully non-perturbative classical lattice simulation of kinetic fragmentation in the context of axion dark matter. To this end, we limit our analysis to the QCD axion as that is the most well motivated scenario from the perspective of the standard model, and also the most compelling parameter space within reach of axion haloscopes. The parameter space we studied is highlighted in blue in Fig.~\ref{fig:experiments}. We narrowed our analysis to this range because for smaller masses, the QCD axion does not fragment and therefore the evolution is well captured by a linear approximation. One the other hand, the upper limit is due to limits in our computational resources. Our goal is to confront the approximate results of linear theory and homogeneous backreaction against the non-perturbative lattice simulations and get a quantitative grasp of how non-perturbative dynamics change the conclusions summarized in the previous paragraph. Our lattice simulations were performed using an adapted version of {\tt ${\mathcal C}$osmo${\mathcal L}$attice} \cite{Figueroa:2020rrl,Figueroa:2021yhd}.

This paper is organized as follows. In Sec~\ref{Sec:kinetic-misalignment} we introduce the analytic results of the standard kinetic misalignment and establish the system of equations to be solved numerically to study kinetic fragmentation. In Sec.~\ref{Sec:kinetic-fragmentation} we present two numerical strategies for simulating kinetic fragmentation. One which uses linear perturbation theory with homogeneous backreaction, and therefore is analogous to previous simulations in the literature, and the fully non-perturbative lattice simulation which is the main novelty of this work. In Sec~\ref{Sec:comparison} we compare the results of the two levels of approximation with each other and give a qualitative and quantitative account of the differences as well as the implications in the grand scheme of kinetic fragmentation. We conclude in Sec.~\ref{sec:conclusions} and relegate some subtle points deemed too long to be in the main text in two Appendices~\ref{app:initial-conditions} and~\ref{app:homogeneous-backreaction}.

%
\section{Kinetic misalignment of the QCD axion}
\label{Sec:kinetic-misalignment}
%
%
\subsection{Axion dark matter abundance and trapping temperature}
\label{Sec:axion-abundance}
%
%
We start by reviewing the kinetic misalignment mechanism. In order to guide our numerical analysis, it is helpful to first derive some analytical results. These results are particularly valid at early times, well before the onset of fragmentation. We will  only very briefly comment on the origin of these results and instead refer the reader to \cite{Eroncel:2022vjg}, a thorough reference on the topic. The Lagrangian that governs the physics of the QCD axion takes the form
\begin{equation}
    {\cal L}=-\frac{f^2}{2}g^{\mu\nu}\partial_\mu \theta\,\partial_\nu\theta-V(\theta)=-\frac{f^2}{2}g^{\mu\nu}\partial_\mu \theta\,\partial_\nu\theta-m^2(T)f^2\left[1-\cos(\theta)\right]\,.
\end{equation}
where $\theta\equiv\phi(t,\vec{x})/f$ is the axion misalignment angle whose space-time dependence will typically be omitted for brevity. The mass of the axion is temperature dependent and increases quite rapidly until about the temperature of the QCD phase transition, after which it becomes constant. We will approximate the temperature dependence in the same manner as \cite{Eroncel:2022vjg}, which in turn reproduces well the lattice QCD results  \cite{Borsanyi:2016ksw}. The mass at high temperatures can be written as
\begin{equation}
    m^2(T)=m_0^2\,\times\,\begin{cases}
  (T/T_{\rm c})^{-\gamma} & \text{for}\;\; T > T_{\rm c} \\
  1 & \text{for}\;\; T\leq  T_{\rm c}\,,
  \end{cases}
  \label{eq:mass}
\end{equation}
where $T_{\rm c} \equiv 2.12\,\Lambda_{{\rm b},0}$, $\gamma=8.16$ and $m_0^2 f^2= \Lambda_{{\rm b},0}^4\equiv \left(75.6 \;{\rm MeV}\right)^4$  is the zero temperature barrier height of the axion potential. 

We split the axion field into a time-dependent background and inhomogeneous fluctuations $\theta\equiv \Theta(t)+\delta\theta (t,\vec{x})$. Fragmentation occurs due to the growth of fluctuations caused by non-linearities and their effect on the background dynamics. It is nevertheless instructive to study the physics of the kinetic misalignment by temporarily  disregarding the effect of fluctuations. In the absence of fragmentation, there are three distinct eras in the evolution of the kinetic misalignment mechanism that are noteworthy. The early phase, during which the kinetic energy of the axion dominates over the potential energy. During this time the axion rapidly sheds its kinetic energy,  as $\rho_\Theta\propto a^{^{-6}}$. The next to follow is the ``trapping" period when the axion potential becomes comparable to its kinetic energy and the axion is trapped by one of the potential minima.  Finally, we have the last stage of the evolution: after the mass has acquired its zero temperature value, the axion decays as matter, $\rho_\Theta\propto a^{-3}$, oscillating around the potential minimum. 

Crucial ingredients for studying the kinetic misalignment numerically are the final axion abundance $\Omega_{{\rm a},0}h^2$ and the axion trapping temperature $T_*$. In general, we will reserve the $*$ notation for quantities that are evaluated at the moment of trapping. The most efficient way to compute the axion dark matter abundance is to use the Wentzel–Kramers–Brillouin (WKB) approximation: it allows for the derivation of a well-defined adiabatic invariant which is conserved to very high degree during the three phases described above. In order to facilitate comparison with previous literature, we will use the notation of \cite{Eroncel:2022vjg}. Since we need to associate the temperature of trapping with the final dark matter abundance, it is enough to evaluate the adiabatic invariant $J$ at the moment of trapping and at present times:
\begin{equation}
    J_*\simeq 8 m_* f^2 a_*^3 \;\;\;;\;\;\;J_0\simeq\pi \frac{\rho_{\Theta,0} a^3}{m_0}\; .
\end{equation}
Enforcing the conservation of the adiabatic invariant as well as entropy conservation, it is possible to derive a relationship between the present axion abundance and the trapping temperature:
\begin{equation}
    \rho_{\Theta,0}=\frac{8 \Lambda_{{\rm b},0}^4}{\pi}\left(\frac{T_0}{T_*}\right)^3\left(\frac{T_*}{T_{\rm c}}\right)^{-\gamma/2}\frac{g_{s,0}}{g_{s,*}}\;.
    \label{eq:zero-energy}
\end{equation}
The right hand of Eq.~(\ref{eq:zero-energy})  depends solely on the trapping temperature $T_*$ and various constants, so that there is a unique trapping temperature that yields the observed dark matter abundance for a QCD axion of any mass. This is a well-known result \cite{Papageorgiou:2022prc,Eroncel:2022vjg}. The  temperature is found to be approximately
\begin{equation}
    T_*^{\rm QCD}\simeq\left(1.18 \;{\rm GeV}\right) \left(\frac{\Omega_{\Theta,0}h^2}{\Omega_{\rm DM} h^2}\right)^{-0.141}\;.
    \label{eq:trapping-QCD}
\end{equation}
%
%
\subsection{QCD axion parametrization and initial conditions}
\label{Sec:parametrization}
%
%
Once the trapping temperature is uniquely determined, different values of the axion mass should correspond to different initial kinetic energies. In the kinetic misalignment scenario the mass of the axion can be as low as  $m_0\sim 10^{-5}\;{\rm eV}$: such an axion would be trapped at temperature $T_*^{\rm QCD}$ with vanishing initial kinetic energy. The axion would start its oscillations around the minimum of the potential when the well-known condition for the standard misalignment mechanism is met, namely $m_*\simeq 3 H_*$. As the mass of the axion increases, so should the initial kinetic energy in order to ensure the correct trapping temperature $T_*^{\rm QCD}$.  It is convenient to use an auxiliary variable, $n$, to parametrize the mass of the axion:
\begin{equation}
    n\equiv \frac{m_*}{H_*}\;,
    \label{eq:ndef}
\end{equation}
with $n=3$ as the lowest  value corresponding to the conventional misalignment mechanism, and $n\simeq 17000$ as the highest value given that masses larger than about $m_0\simeq 10^{-1}\;{\rm eV}$ are experimentally ruled out  \cite{AxionLimits}. 
Of course, there is an exact one-to-one correspondence between $n$ and the zero temperature mass of the axion. In practice, we will limit our analysis to the interval  $50 \leq n \leq 2000$ as such domain includes most of the mass range accessible by future haloscopes (see Fig.~\ref{fig:experiments}) and covers a range of axions that fragment both before as well as after trapping. Table~\ref{tab:f_and_m} includes the values of the axion decay constant $f$, zero temperature mass $m_0$ and trapping mass $m_*$ of the cases studied in this work.

\begin{table}[h]
\centering
{\renewcommand{\arraystretch}{1.25}
\begin{tabular}{|c|c|c|c|c|}
\hline
$n$ & $f\ (\text{GeV})$ & $m_0\ (\text{eV})$ & $m_*\ (\text{eV})$\\
\hline
50 & $2.0231\cdot10^{10}$ & $2.8251\cdot10^{-4}$& $8.2864\cdot10^{-8} $  \\
100 & $1.0115\cdot10^{10}$ & $5.6502\cdot10^{-4}$& $1.6572\cdot10^{-7} $  \\
200 & $5.0576\cdot10^{9}$ & $1.1300\cdot10^{-3}$& $3.3146\cdot10^{-7} $\\
300 & $3.3718\cdot10^{9}$ & $1.6950\cdot10^{-3}$& $4.9719\cdot10^{-7} $  \\
400 & $2.5288\cdot10^{9}$ & $2.2601\cdot10^{-3}$& $6.6291\cdot10^{-7} $  \\
500 & $2.0231\cdot10^{9}$ & $2.8251\cdot10^{-3}$& $8.2864\cdot10^{-7} $ \\
1000 & $1.0115\cdot10^{9}$ & $5.6502\cdot10^{-3}$& $1.6572\cdot10^{-6} $  \\
1500 & $6.7435\cdot10^{8}$ & $8.4753\cdot10^{-3}$& $2.4859\cdot10^{-6} $  \\
2000 & $5.0576\cdot10^{8}$ & $1.1300\cdot10^{-2}$& $3.3146\cdot10^{-6} $ \\
\hline
\end{tabular}
\caption{Parameters of the cases studied in this work, where $f$ corresponds to the axion decay constant, $m_0$ is the zero temperature mass and $m_*$ is the trapping mass.} }
\label{tab:f_and_m}
\end{table}

The calculation we are aiming to do boils down to evolving the equation of motion for the axion in an expanding background. We do not evolve the background geometry because the total energy of the axion is a negligible fraction of the total energy of the universe in the regime under consideration here. The axion equation of motion takes the form
\begin{equation}
    \ddot{\theta}+3H\dot{\theta}-\frac{1}{a^2}\nabla^2\theta+m(T)^2\sin\left(\theta\right)=0
    \label{eq:system}
\end{equation}
We use the subscript ``$\rm ini$" to denote variables evaluated at the start of our simulation. The scale factor in radiation domination is $a=(t/t_{\rm ini})^{1/2}=(T_{\rm ini}/T)$ which implies that we are disregarding the change of the effective degrees of freedom during the evolution, and employ their values at the moment of trapping. Our initial conditions are set at an early time when the kinetic energy of the axion is much greater than the potential energy. Let us define a parameter as the ratio between the initial kinetic and potential energies
\begin{equation}
    c\equiv \frac{\frac{1}{2}\dot{\Theta}_{\rm ini}^2}{m(T_{\rm ini})^2\left[1-\cos\left(\Theta_{\rm ini}\right)\right]} \; .
    \label{eq:kinini}
\end{equation}
Since at early times the potential accounts for a very small fraction of the total energy of the universe, we may arbitrarily choose  an initial misalignment angle near the top of the potential $\Theta_{\rm ini}=\pi$\footnote{It turns out the solution does depend mildly on the exact value of the initial misalignment angle. We shall comment further on this matter in section \ref{Sec:fragmentation-moment}.}. The kinetic energy at early times scales approximately as kination $\dot{\Theta}^2\propto a^{-6}$. Considering Eq.~(\ref{eq:mass}), the scaling of the potential energy  is governed by the evolution of the mass $m(T)\propto a^\gamma$. We can then relate the trapping temperature to the initial temperature,
\begin{equation}
    \left(\frac{T_{\rm ini}}{T_{*}^{\rm QCD}}\right)^{6+\gamma}=c\;.
    \label{eq:Tini}
\end{equation}

Eqs. (\ref{eq:kinini}) and (\ref{eq:Tini}) can be viewed as two independent equations with two unknowns, namely $T_{\rm ini}$ and $\dot{\Theta}_{\rm ini}$. Solving for these two variables fully fixes the initial conditions for the background. In practice, we find that the values around  $c=1000$, which corresponds to $T_{\rm ini} = 1.917\  \text{GeV}$, to be in the ``sweet spot'': sufficiently far from trapping so that the initial assumption of kinetic domination remains valid, and not too computationally-expensive. We have performed checks with $c=10^5$ and $c=10^6$, obtaining indistinguishable results.

Our simulation evolves from an initial value for the scale factor of $a_{\rm ini}=1$, to a final value of $a_{\rm end}=2.5$. The initial conditions for the fluctuations are non-trivial and we address them in detail in Appendix \ref{app:initial-conditions}. In the next section, we tackle the system in Eq.~(\ref{eq:system}) at increasing levels of sophistication. We begin under the assumption of homogeneous backreaction and later present a fully non-perturbative lattice simulation.
%
%
\section{Backreaction and kinetic fragmentation of the QCD axion}
\label{Sec:kinetic-fragmentation}
%
%
It is well-known that an axion zero mode rolling rapidly over many potential maxima will experience tachyonic instability and parametric resonance \cite{Fonseca:2019ypl}. As a result, fluctuations may be sufficiently enhanced to significantly backreact on the axion zero mode, altering its dynamics and potentially exceeding its energy, i.e. causing fragmentation. Understanding backreaction dynamics is therefore crucial. This task has so far been addressed either by assuming homogeneous backreaction or through analytical methods within perturbation theory \cite{Fonseca:2019ypl,Eroncel:2022vjg}. 

In this section we study axion fragmentation by first assuming homogeneous perturbative backreaction and later putting forward fully non-perturbative lattice simulations. The first method is the closest to results already in the literature and so we find it useful both as a reference point and as a first step towards the full lattice study. The comparison between the two approaches will highlight the effects that are missed by  the simplified homogeneous backreaction study. 
 
%
%
\subsection{Homogeneous backreaction approximation}
\label{Sec:homogeneous-backreaction}
%
%
The dynamics of the full system, given in Eq. (\ref{eq:system}), can be decomposed into an equation for the zero mode and an equation for perturbations. We will consider the lowest order backreaction term that is proportional to the third derivative of the potential $V^{(3)}(\theta)$\footnote{The lowest-order backreaction term for a generic scalar is given by: $\frac{1}{2}V^{(3)}(\phi)\int\frac{d^3x}{V_{\rm vol}}\langle\delta\phi(x)^2\rangle$.}.
\begin{align}
 \ddot{\Theta}&+3H\dot{\Theta}+m(T)^2 \sin(\Theta)=\frac{m(T)^2\sin(\Theta)}{2}\int\frac{d^3k}{(2\pi)^3}|\theta_k|^2\; ;\nonumber\\
 \ddot{\theta}_k&+3H\dot{\theta}_k+\left[\frac{k^2}{a^2}+m(T)^2\cos(\Theta)\right]\theta_k=0\; .
 \label{eq:eomnume}
\end{align}
Our Fourier convention is 
\begin{equation}
    \delta \theta(t,\vec{x})=\int\frac{d^3k}{\left(2\pi\right)^{3/2}}\theta_k\,{\rm e}^{-i\vec{k}\cdot\vec{x}}\;,
\end{equation}
where statistical isotropy is assumed, hence the mode functions are dependent only on the amplitude of the comoving momentum $k$. The system of equations under scrutiny is linear in the sense that all perturbations evolve independently of one another. It makes up a set of integro-differential equations whose solution can be computed straightforwardly with the discretization method. This method requires discretizing the comoving momentum and then reconstructing the integral (\ref{eq:eomnume}) at every moment in time with the trapezoid rule. As long as the discretization step is small enough, the trapezoid rule reproduces the continuous integral to a very high degree. In practice, we choose a momentum window so that modes relevant for backreaction are taken into account at all times and discretize the momentum so that the step is constant in $\log k$ space. More precisely, we select
\begin{equation}
    k_{\rm min}=30\, a_*H_*\;\;\;,\;\;\;k_{\rm max}\simeq 2.25 \,k_{\rm thr,*}\;\;\; {\rm and}\;\;\; k[i]\equiv k_{\rm min}\left(\frac{k_{\rm max}}{k_{\rm min}}\right)^{\frac{i-1}{i_{\rm max}-1}}\; ;
\end{equation}
where $i_{\rm max}=300$ is the total number of modes used and
\begin{equation}
    k_{\rm thr,*}\equiv a_* m_*\left(\frac{4}{\pi}a_*^{-\frac{1}{2}+\frac{\gamma}{4}}\right)^{1/2}\;
\end{equation}
is the upper limit of the instability in the scalar perturbations at the moment of fragmentation as derived by the Floquet analysis of \cite{Eroncel:2022vjg} (specifically Eq.~(3.21) therein) . The reasoning behind the upper limit is to include all the momenta that are naively expected to be resonantly enhanced by the ``backreactionless'' analysis. On the other hand, the result is less sensitive to the lower limit as long as all modes that are sufficiently enhanced and have an impact on the backreaction are included. As we show below, the use of these criteria in our numerical analysis guarantee at all times a high degree of precision in accounting for backreaction. 

We  shall rewrite the integral on the right hand side of Eq.~(\ref{eq:eomnume}) with the trapezoid rule as follows. For a generic function ${\cal I}(k)$,
\begin{align}
    \int^{k_{\rm max}}_{k_{\rm min}} {\cal I}(k) \,dk \Rightarrow \sum^{i_{\rm max}-1}_{i=1}\Delta\ln k_i \frac{k_i {\cal I}_{k_i}+k_{i+1} {\cal I}_{k_{i+1}}}{2}= \label{trap_1}\\ \frac{\Delta\ln k}{2}\left(k_{\rm min}{\cal I}_{k_{\rm min}}+k_{\rm max}{\cal I}_{k_{\rm max}}+2 \sum^{i_{\rm max}-1}_{i=1}k_i {\cal I}_{k_{\rm i}}\right)\;. \label{trap_2}
\end{align}
The above ``$\Rightarrow$'' arrow signals the application of the trapezoid rule to approximate the continuous integral, while the second equality is valid only when $\ln k_{i+1}-\ln k_i$ is constant, which in our case is true by construction. Finally, the first two terms in Eq.~(\ref{trap_2}) within the parentheses are negligible boundary terms we will therefore disregard. In Appendix \ref{app:homogeneous-backreaction} we include more technical details on simulating the homogeneous backreaction and on some ambiguities in choosing a criterion for fragmentation.

While it is true that the system in Eq. (\ref{eq:eomnume}) is formally valid only up to the moment of fragmentation, we also attempt to simulate the post-fragmentation era by evolving the equations of motion for the perturbations alone with the zero mode assumed to be constant at the bottom of the potential $\Theta=0$. As we will verify later, this approximation is qualitatively correct but, for an accurate picture of what occurs at late times, lattice simulations are necessary. 

\begin{figure}
    \centering
    \includegraphics[width=0.495\linewidth]{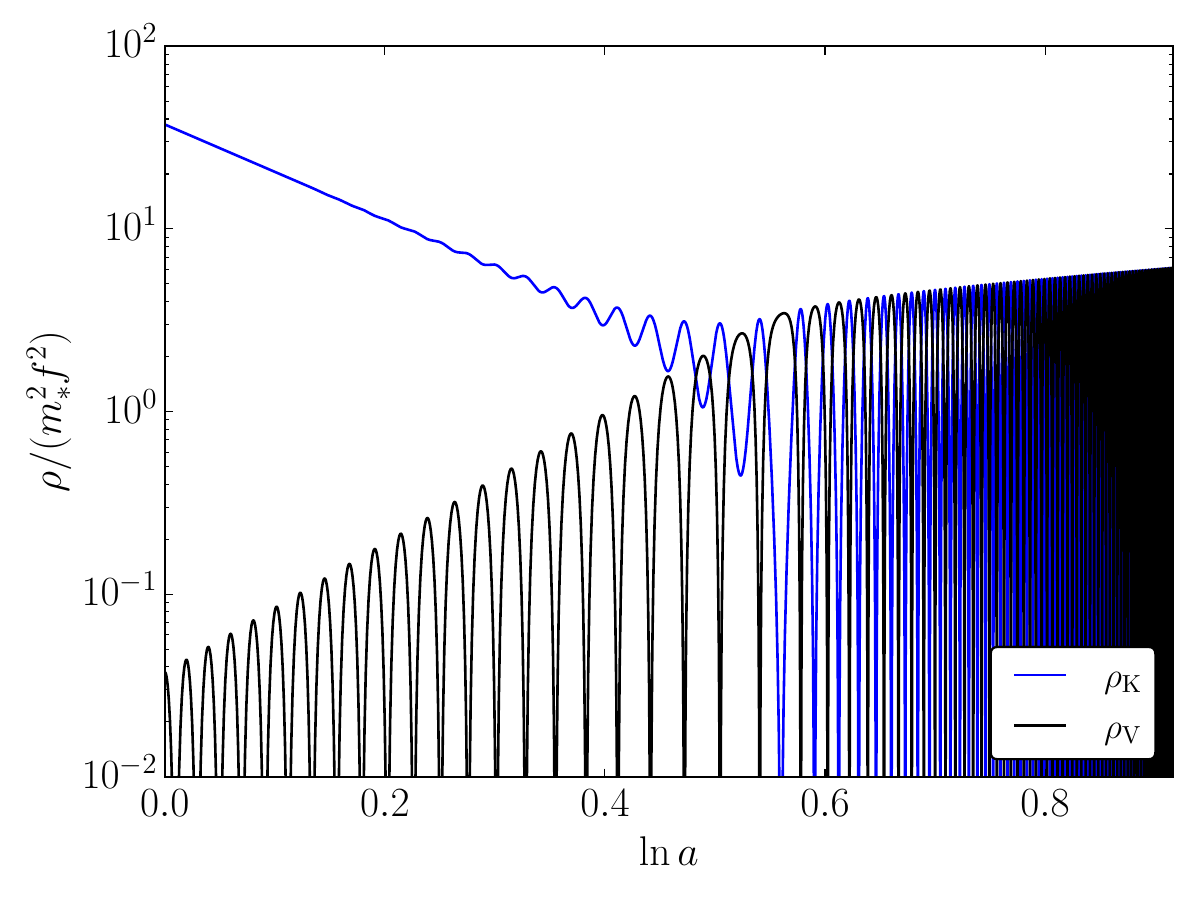}
    \includegraphics[width=0.495\linewidth]{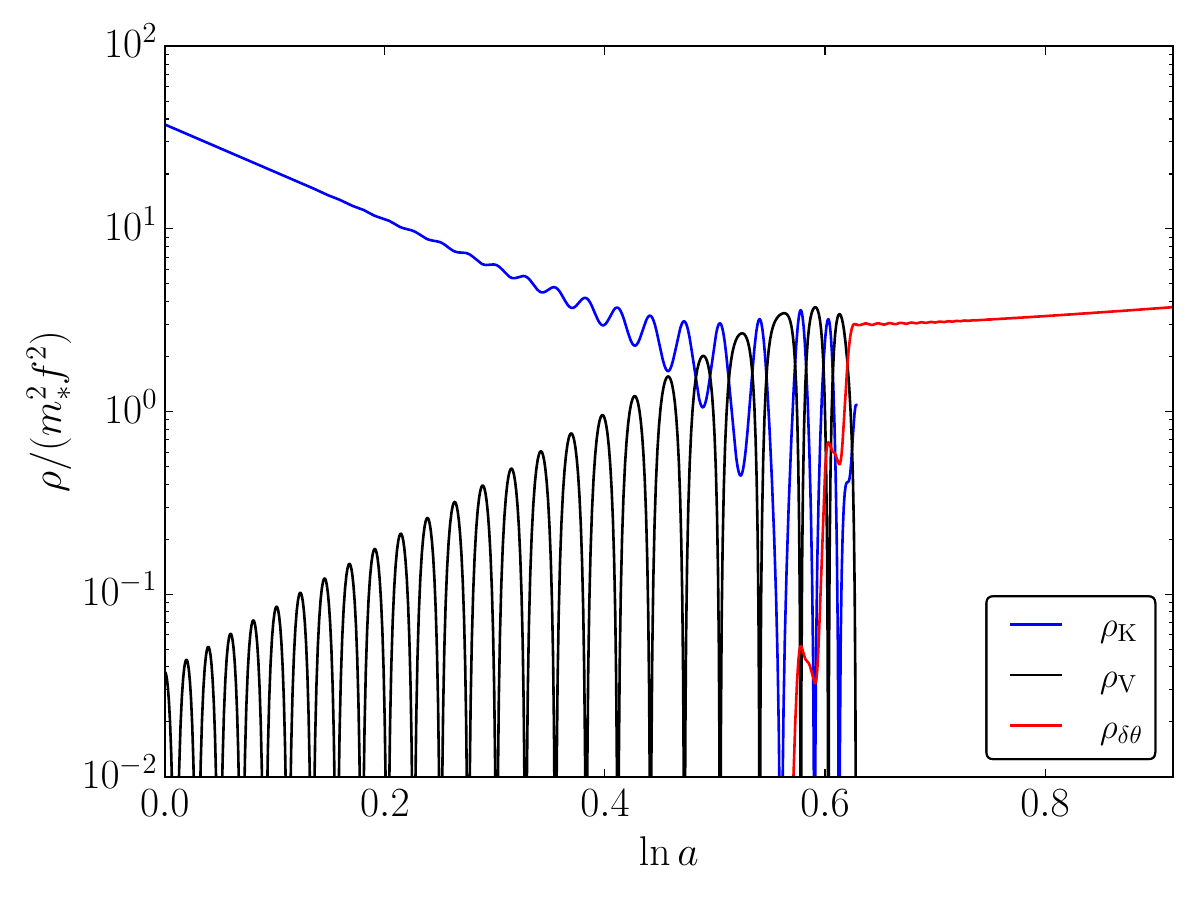}
    \caption{\textit{Left panel}: evolution of the kinetic and potential energy densities of the zero mode. \textit{Right panel}: evolution of the kinetic and potential energy of the zero mode along with the energy of perturbations in red. In this case the backreaction is treated according to the homogeneous approximation outlined in Sec.~\ref{Sec:homogeneous-backreaction}.}
    \label{fig:hom-bck}
\end{figure}

The final results of the homogeneous backreaction analysis are displayed in Fig.~\ref{fig:hom-bck}. We choose $n=100$ to provide an illustrative sample of the dynamics emerging in the homogeneous backreaction approximation. On the left panel of Fig.~\ref{fig:hom-bck}, the evolution of the energy density of the zero mode is displayed without including backreaction effects. The right panel includes instead  the effects of backreaction and also shows the post-fragmentation evolution of the perturbations in the way we outlined above. One can see that at early times, the kinetic energy is, as expected, much greater than the potential energy of the axion while the energy of perturbations is negligible. At some moment after trapping the energy of perturbations increases rapidly and starts affecting the dynamics of the zero mode until the point around $\ln a\simeq 0.62$ when the energy density in perturbations becomes dominant. Fragmentation occurs when all the energy of the axion zero mode is depleted, leaving only the perturbations evolving at late times.
%
%
\subsection{Fully non-perturbative lattice simulations}
\label{Sec:lattice}
%
%
For the non-perturbative analysis of the dynamics we use lattice simulation. For this end, we employ the basic toolkit of the {\tt ${\mathcal C}$osmo${\mathcal L}$attice} library \cite{Figueroa:2021yhd} for scalar fields, where the equations of motion are evolved in grids with $N$ points per dimension, $\delta x$ spatial lattice spacing, and with periodic boundary conditions. The background geometry is fixed and assumed to expand at a constant rate, corresponding to radiation domination. In this section, we briefly outline the key aspects of the numerical setup and refer the reader to \cite{Figueroa:2020rrl} for further details.

We use the conformal time, $d\tau/d=a^{-1}dt/d=(...)'$, version of Eq.~(\ref{eq:system}) redefined as
\begin{equation}
    \pi'_{\theta}-a^2\nabla^2\theta+a^4m^{2}(H)\sin\theta=0\,,
\end{equation}
where $\pi_{\theta}=a^2\theta'$ and we use the Hubble parameter dependent version of Eq.~(\ref{eq:mass}) with $m^2(H)=m^2_0(H/H_c)^{-\gamma/2}$. This equation is discretized and evolved using the standard \textit{Velocity-Verlet} time integrator of second order (VV2), that follows the kick-drift-kick scheme:
\begin{equation}
\begin{split}
        (\pi_{\theta})_{+\frac{\hat{0}}{2}}&=\pi_{\theta}+\frac{\delta \tau}{2}\mathcal{K}[\theta,a,H]\;,\\
        \theta_{+\hat{0}}&=\theta+\delta \tau (\pi_{\theta})_{+\frac{\hat{0}}{2}}a^{-2}_{+\frac{\hat{0}}{2}}\;,\\
        (\pi_{\theta})_{+\hat{0}}&=(\pi_{\theta})_{+\frac{\hat{0}}{2}}+\frac{\delta \tau}{2}\mathcal{K}[\theta_{+\hat{0}},a_{+\hat{0}},H_{+\hat{0}}]\;,
\end{split}
\label{eq:VV2}
\end{equation}
where the discrete kernels are defined  as
\begin{equation}
    \mathcal{K}[\theta,a,H]=\pi^{'}_{\theta}=a^2\sum_i\Delta^{+}_{i}\Delta^{-}_{i}\theta-a^4m^{2}(H)\sin\theta\,.
\end{equation}
As customary in canonical scalar theories, the scalar field $\theta(\mathbf{n})$ and its conjugate momentum $\pi_\theta(\mathbf{n})$ are both defined on the sites $\mathbf{n}$ of the lattice, while forward and backward spatial derivatives, $\Delta^\pm_{i}\theta(\mathbf{n}+\hat{\imath}/2)=[\pm \theta(\mathbf{n}\pm\hat{\imath})\mp \theta (\mathbf{n}))]/\delta x$, are implemented halfway between neighbour sites. The subscript $+\hat{0}$ denotes a shift of the discrete field by one time step $\delta \tau$. Note that we employ the VV2 scheme in Eq.~(\ref{eq:VV2}), which requires a  $+\hat{0}/2$ half-step update of $\pi_\theta$ to evolve the system.

This scheme reproduces the continuum dynamics up to order $\mathcal{O}(\delta \tau^2,\delta x^2)$. We have cross-checked the results using higher-order schemes (VV3 and VV4), finding no differences in the physical observables relevant to our study, which confirms our framework provides sufficient accuracy for the problem at hand. Simulations are performed in conformal time, which allows for smaller effective time steps $\delta t$ as the system approaches the trapping regime. In this sense, the temporal spacing $\delta\tau = 0.001/m_{*}$ of each case is adequate to capture the relevant features of the evolution, and the spatial resolution $\delta x$ is chosen accordingly to capture the physically relevant scales for each $n$. The values of $N$, $k_\mathrm{IR}/m_{*}$ and $k_\mathrm{UV}/m_{*}$ used in each simulation can be found in Table~\ref{tab:Latt_parameters}.

\begin{table}[h]
\centering
{\renewcommand{\arraystretch}{1.25}
\begin{tabular}{|c|c|c|c|}
\hline
$n$ & $N$ & $k_\mathrm{IR}/m_{*}$ & $k_\mathrm{UV}/m_{*}$ \\
\hline
50 & 672 & 0.2 & 124.71 \\
100 & 360 & 0.2 & 62.35 \\
200 & 360 & 0.2 & 62.35 \\
300 & 360 & 0.2 & 62.35 \\
400 & 360 & 0.2 & 62.35 \\
500 & 360 & 0.2 & 62.35 \\
1000 & 2016 & 0.1 & 174.59 \\
1500 & 2016 & 0.1 & 174.59 \\
2000 & 2016 & 0.1 & 174.59 \\
\hline
\end{tabular}
\caption{Lattice parameters used in the simulations, see text for explanations.}
\label{tab:Latt_parameters}}
\end{table}

The initial conditions are set following the expressions in Appendix \ref{app:initial-conditions}. In particular, we add fluctuations to the initial homogeneous background solution using Eq.~\ref{eq:initial-conditions} (without the cosine) and its time derivative for $\theta$ and $\pi_\theta$, respectively, and follow the standard procedure described in \cite{Figueroa:2021yhd,Figueroa:2020rrl} for setting initial conditions of the scalar field as Gaussian random variables.

Finally, by solving the discretized equations of motion at each lattice point, we obtain full access to the complete field distribution across the entire simulation volume. Nevertheless, a prescription to extract the energy density stored only in fluctuations is necessary. We follow the definition in \cite{Adshead:2019lbr} and define it as

\begin{equation}
    \rho_{\delta\theta} \equiv \rho_{\rm tot} - \rho_{\Theta}\, ,
    \label{eq:fluctuation-energy}
\end{equation}
where $\rho_{\rm tot}$ is the total averaged energy density and $\rho_{\Theta}$ is the energy density of the zero mode. On the lattice this takes the form
\begin{equation}
    \rho_{\Theta} = \rho_{\bar{\theta}} =  \frac{1}{2a^2}{\bar{\theta}'} + V(\bar{\theta})\, ,
\end{equation}
where $\bar{\theta}$ is the mean value of the field and $\bar{\theta}' = (\bar{\theta}_{+\hat{0}} - \bar{\theta}_{-\hat{0}})/(2\delta\tau)$.

%
%
\section{Homogeneous vs Inhomogeneous Dynamics}
\label{Sec:comparison}
%
%
Having at our disposal the tools to simulate  fragmentation both in the homogeneous backreaction approximation and on the lattice, we performed the analysis for a series of representative values of the parameter $n$ defined in Eq.~(\ref{eq:ndef}). We devote this section to a comparison of the dynamics in the two approximations. Naturally, the non-perturbative lattice analysis is the benchmark with respect to which the perturbative and homogeneous backreaction approximation is to be evaluated. 
%
%
\subsection{Comparison of the energy densities}
\label{Sec:comparison-energy}
%
%
\begin{figure}[h!]
    \centering
    \includegraphics[width=0.495\linewidth]{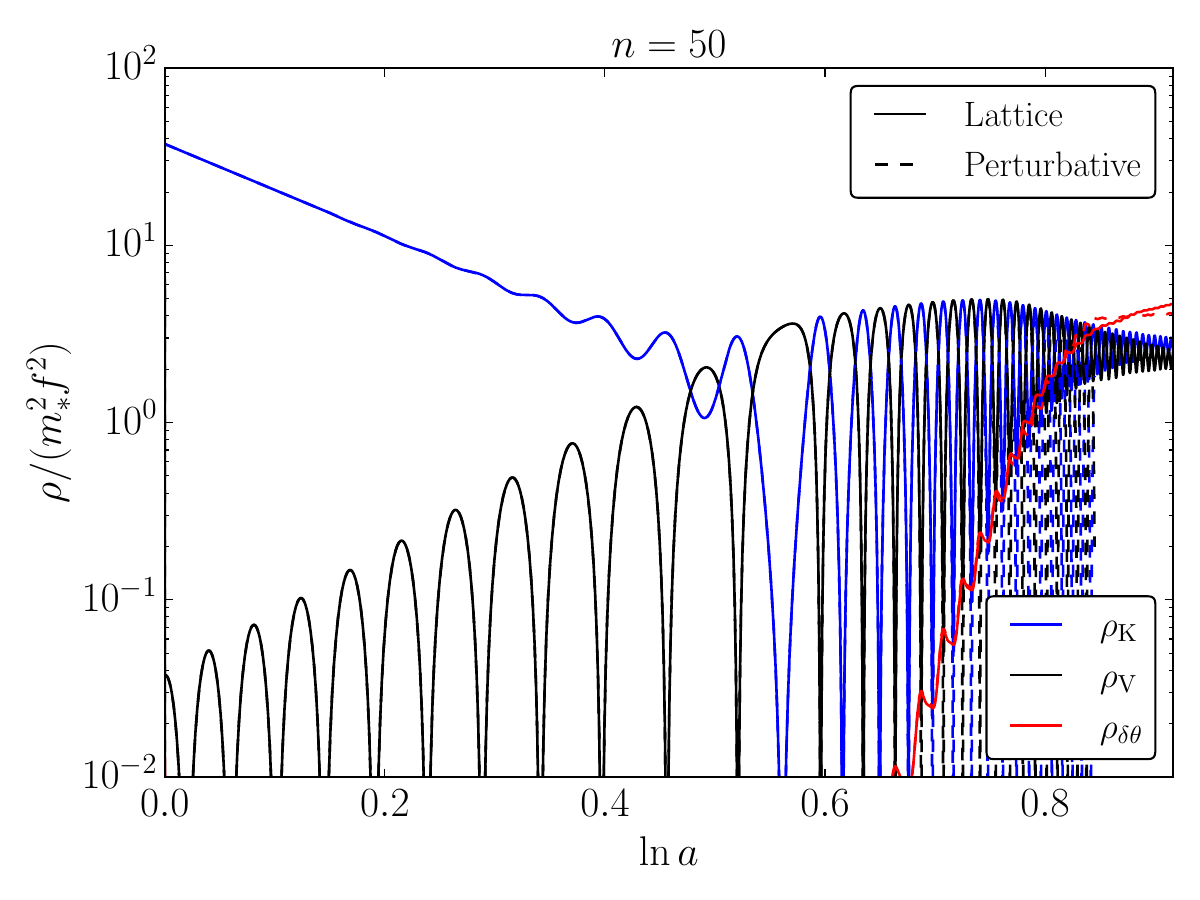}
    \includegraphics[width=0.495\linewidth]{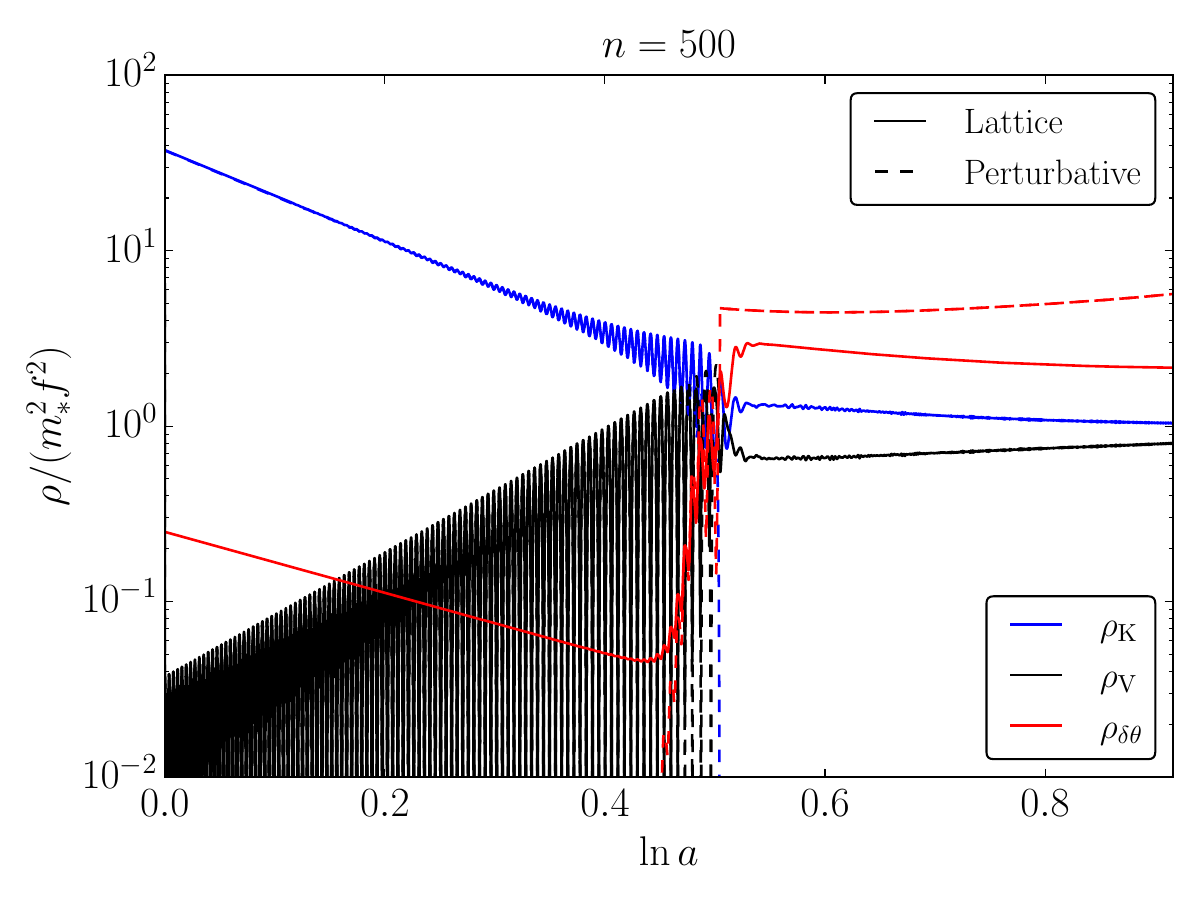}\\
    \includegraphics[width=0.495\linewidth]{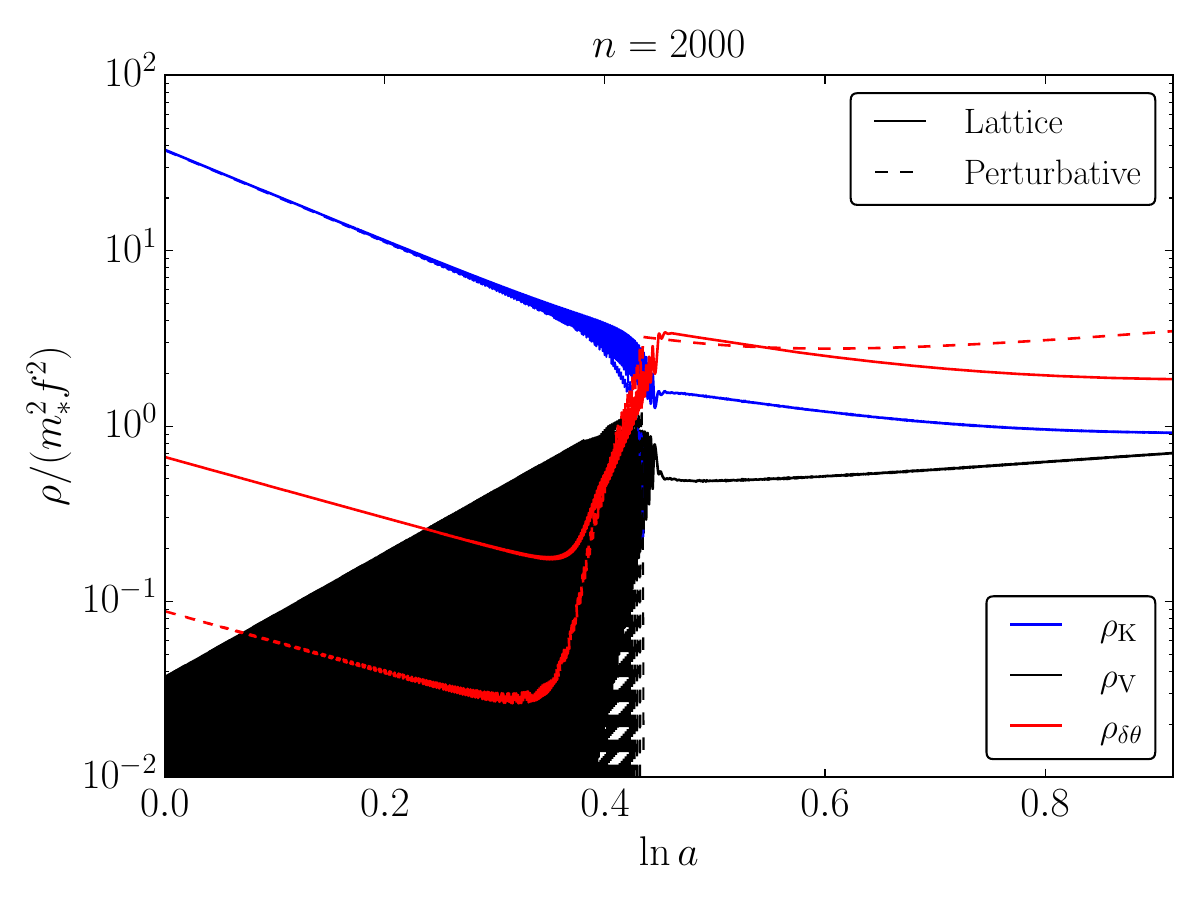}\\
    \caption{The evolution of the rescaled energy densities in the lattice analysis is displayed in solid lines for various values of $n$. Dashed lines show the evolution of the energy density of fluctuations in the homogeneous backreaction approximation.}
    \label{fig:rho_comp}
\end{figure}
We start by comparing, as a function of time, the kinetic, potential, and gradient energies from the lattice analysis to the energy in fluctuations in the homogeneous approximation. In Fig.~\ref{fig:rho_comp} the solid lines show the results for the lattice analysis \textit{vs} the energy of fluctuations in the homogeneous case in dashed red lines. As expected, at very early times, when the system is well approximated by perturbation theory, there is very good agreement between the lattice and the homogeneous backreaction results for the energy of fluctuations. Additionally, it is easy to see that the moment of fragmentation, defined as the moment when the energy in fluctuations is more than $90\%$ of the total energy of the system, occurs at similar times in both cases. It is also possible to observe that at late times the energy of fluctuations in the homogeneous scheme is larger than the total energy found in the lattice approach. This effect can be understood by turning our attention to the spectrum of fluctuations as a function of time. 
%
%
\subsection{Comparison of the spectra}
\label{Sec:comparison-spectra}
%
%

\begin{figure}
    \centering
    \includegraphics[width=0.85\linewidth]{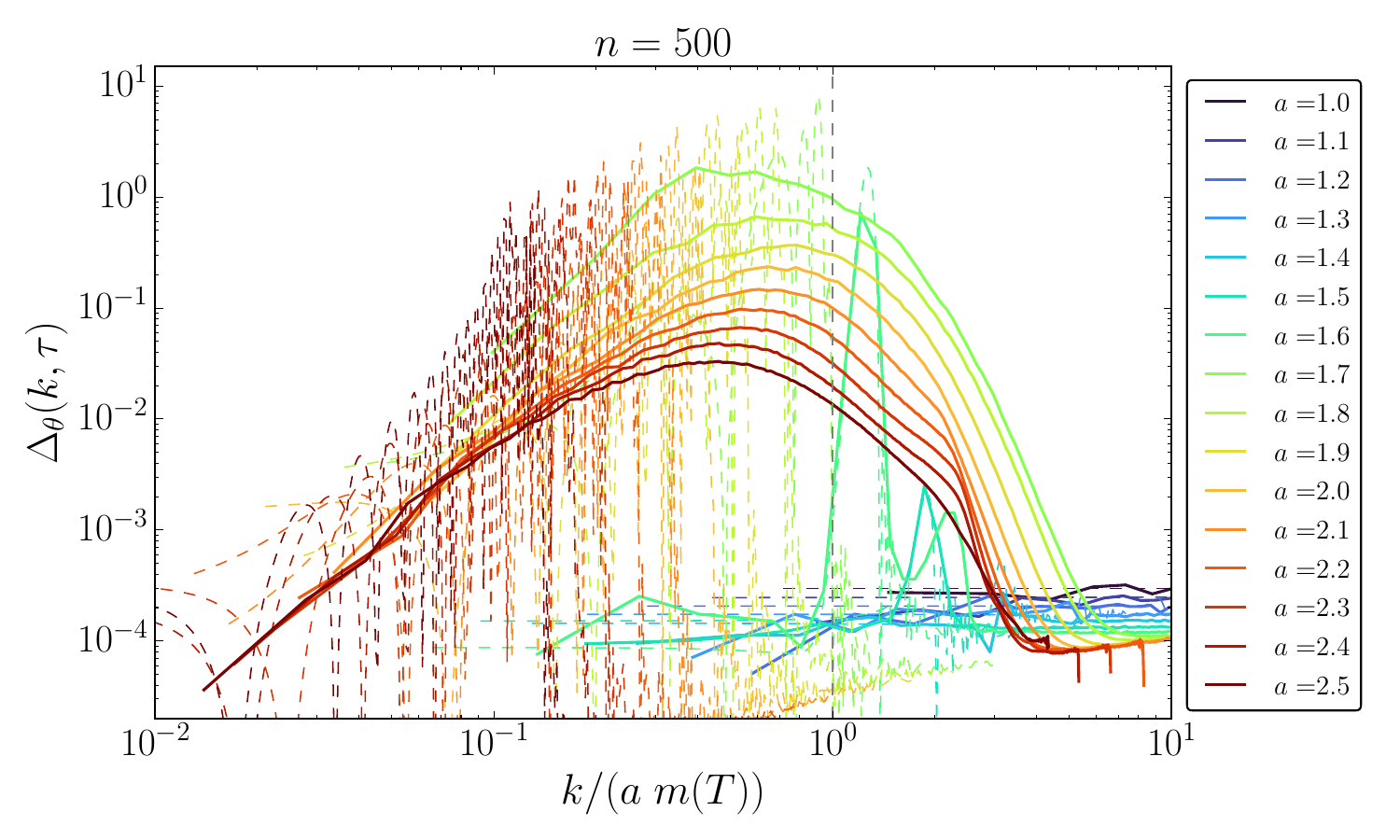}
    \caption{Evolution of the Power Spectrum fo the scalar field for $n=500$, from the beginning of the simulations, $a=1.0$ to $a=2.5$, with steps of $\Delta a = 0.1$. Solid lines correspond to lattice simulations and dashed lines to homogeneous backreaction.}
    \label{fig:PS_n500_Evo}
\end{figure}

In order to understand the distribution of power in fluctuations and its evolution through time, we plot the power spectrum defined as 
\begin{equation}
    \Delta_\theta(k,\tau)\equiv \frac{k^3}{2\pi^2} |\theta_k(\tau)|^2\, ,
    \label{eq:powerspectrum}
\end{equation}
starting from the beginning of our simulation $a_{\rm ini}=1$ until the end $a_{\rm end}=2.5$ with a step $\Delta a=0.1$. On the lattice we use the {\tt Type~II-Version~1} power spectrum, as defined in \cite{CLPS}. We select $n=500$ as a typical representative value and superimpose the spectra of the homogeneous backreaction (in dashed lines) with the ones from the lattice (solid lines). Our results are displayed in Fig.~\ref{fig:PS_n500_Evo}. The horizontal axis of the figure has been chosen to be $k/(a \;m(T))$ strategically so that $k/(a \;m(T))=1$ (displayed as a grey, dashed, vertical line) indicates the transition below which modes are non-relativistic. The dispersion relation of the perturbations can be read off Eq.~(\ref{eq:eomnume}),
\begin{equation}
    \omega^2\equiv\left(\frac{k^2}{a^2}+m(T)^2\cos(\Theta)\right)\;.
\end{equation}
Modes can then be separated as non-relativistic ones, for which $\omega_{\rm nr}\simeq m(T)$, and relativistic ones for which $\omega_r\simeq \frac{k}{a}$. We also used the fact that after trapping the zero mode typically settles down rather quickly to the minimum of the potential, $\Theta_{\rm late}\sim 0+2\pi k$. If most of the power is concentrated in non-relativistic modes, then one expects an evolution of the energy at late times similar to that of the background, which typically shows an increase at late times as in the left panel\footnote{This increase is expected; it is a consequence of the rapidly increasing potential barriers of the QCD axion with decreasing temperature.} of Fig.~\ref{fig:hom-bck}. On the other hand, if at late times there is a significant portion of the energy in (near) relativistic modes, the total energy in fluctuations will lag behind the homogeneous mode. 

Turning our attention to Fig.~\ref{fig:PS_n500_Evo}, it is possible to observe that until $a=1.3$ there is no noticeable enhancement of the fluctuations. For the next three steps in $a$ value there is instead a very narrow band of fluctuations that is quickly and significantly enhanced, which  coincides with the sudden growth exhibited by the energy in fluctuations in Fig.~\ref{fig:rho_comp}. For these three values the result of the homogeneous backreaction and the lattice analysis are in very good agreement. Such agreement is most noticeable for $a=1.6$ where there is a clear narrow peak for modes that are just beyond the relativistic threshold. the peak is well described by both the homogeneous and lattice analyses. After that moment, fragmentation occurs and as a result the two analyses differ drastically. In the homogeneous backreaction case, there is no possible way for the modes to exchange energy, instead all modes evolve entirely independently. As a result the peak in fluctuations remains very narrow and with really high occupation numbers. This peak drifts to the non-relativistic limit and, for this reason,  there is an increase in the slope of the energy of fluctuations as these non-relativistic modes evolve in a way that resembles the late time limit of the zero-mode analysis. On the other hand, in the lattice analysis the process of fragmentation leads to a great redistribution of energy among the modes resulting,  at late times, in a very wide spectrum with much lower occupation number. The spectrum is much wider with a significant fraction occupying the relativistic or near-relativistic limit at late times.  The overall energy in fluctuations features then a lower slope than both the zero-mode and homogeneous backreaction analyses.

\begin{figure}
    \centering
    \includegraphics[width=0.85\linewidth]{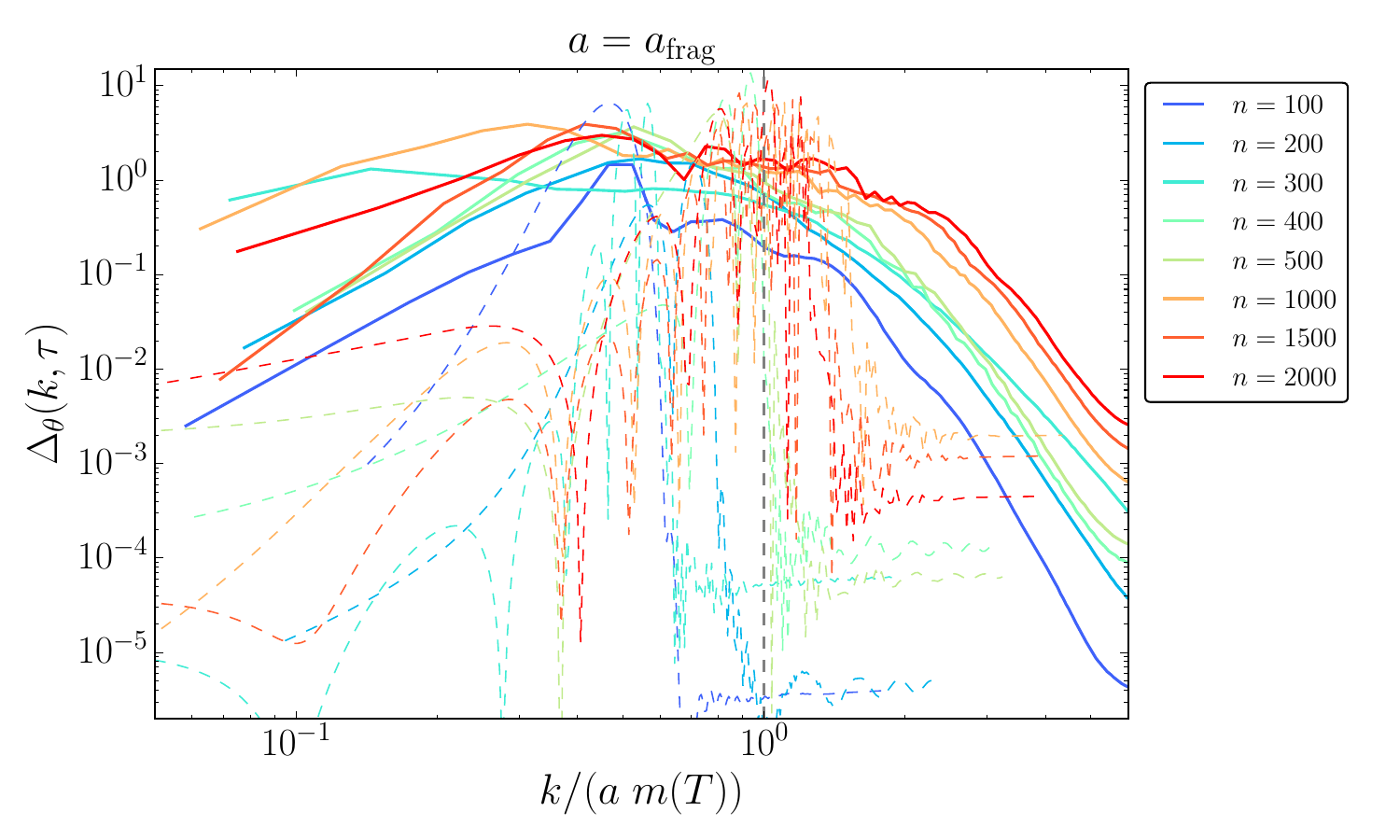}
    \includegraphics[width=0.85\linewidth]{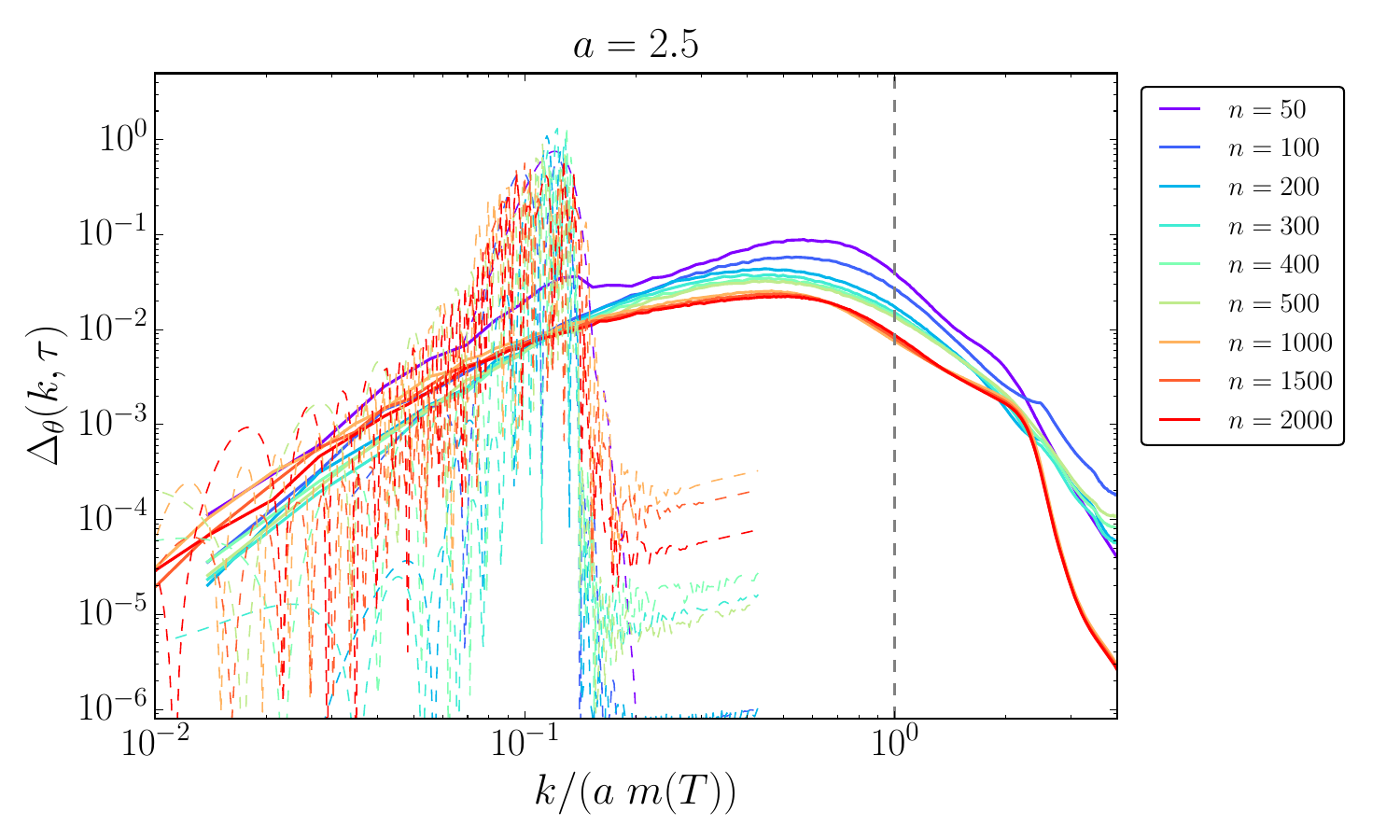}
    \caption{ \textit{Upper panel}: Power spectrum of the scalar field at the moment of fragmentation, defined according Eq.~(\ref{eq:powerspectrum}), for all $n$'s. Solid lines correspond to lattice simulations and dashed lines to homogeneous backreaction. Note that $n=50$ is not included, as it does not completely fragment. \textit{Lower panel}: Equivalent to the left panel for $a=2.5$, i.e. at the end of the evolution.}
    \label{fig:PS_all}
\end{figure}
So far the description of the spectrum  has been based on a specific value of $n$, chosen  as a representative example. Indeed, one notices similar qualitative features for all values of $n$ considered in our analysis with the exception of the $n=50$   (this is not surprising given that for $n=50$  fragmentation does not occur by the end of our simulation).

In Fig.~\ref{fig:PS_all} we compare the spectra between the homogeneous backreaction approximation and the lattice analysis for values of $n$ varying from $n=50$ until $n=2000$. The spectra are compared at two specific moments in the evolution, namely at the moment of fragmentation as computed in the lattice analysis and at the latest possible time which by convention corresponds to a scale factor value of $a=2.5$. Note that $n=50$ is not included in the former, as in that case the system does not reach our criterion for fragmentation. It is possible to observe that qualitatively the spectra are similar across different values of $n$ or, equivalently, different masses of the QCD axion. Especially at the end, the spectrum is very wide in the lattice analysis with a significant amount of power at large comoving wavelengths whereas the homogeneous approximation remains narrow and highly peaked. This universal widening and flattening of the spectrum found in the lattice analysis can have a significant impact on observables associated with the kinetic misalignment mechanisms such as mini-halo formation \cite{Eroncel:2022efc}.
%
%
\section{Discussion}
\label{Sec:Discussion}
%
%
Equipped with the insights gleaned from the comparison between the homogeneousapproach and the lattice analysis, we can proceed to investigate how the non-perturbative dynamics captured by the lattice study aaffects some important aspects of axion fragmentation. We focus on (i) the \emph{moment of fragmentation} which is roughly the time when the energy of the zero mode becomes much smaller than that in fluctuations and (ii) the \emph{dilution factor}, a parameter that estimates how much excess (or lack thereof) energy is produced at late times as a result of fragmentation compared to the zero mode analysis of Eq.~(\ref{eq:zero-energy}).
%
%
\subsection{Moment of fragmentation}
\label{Sec:fragmentation-moment}
%
%

%
Given that fragmentation is the process by which the energy of the zero mode of the axion is being traded for the energy in fluctuations, we use the following concrete criterion to define the moment in which the axion fragments
\begin{equation}
    \frac{\rho_{\delta\theta}}{\rho_{\rm tot}}=90\%             \;\;\;\;,\;\;\;({\rm moment\; of \; fragmentation})
    \label{eq:frag-criterion}
\end{equation}
where $\rho_{\rm tot}$ is the total energy of the system and $\rho_{\delta\theta}$ is the energy in fluctuations of the axion defined in Eq.~(\ref{eq:fluctuation-energy}). Reference~\cite{Eroncel:2022efc} places particular emphasis on whether the axion fragments before or after the moment of trapping.. According to the results in section 4.1.3 of the same reference, the QCD axion fragments at the trapping temperature  when $n\simeq900$ and the corresponding axion decay constant is $f_{\rm crit}\simeq 7.3 \times 10^8\;{\rm GeV}$. It is very important to understand how these values may be altered when fragmentation is simulated on the lattice.

\begin{figure}[h!]
    \centering
    \includegraphics[width=0.75\linewidth]{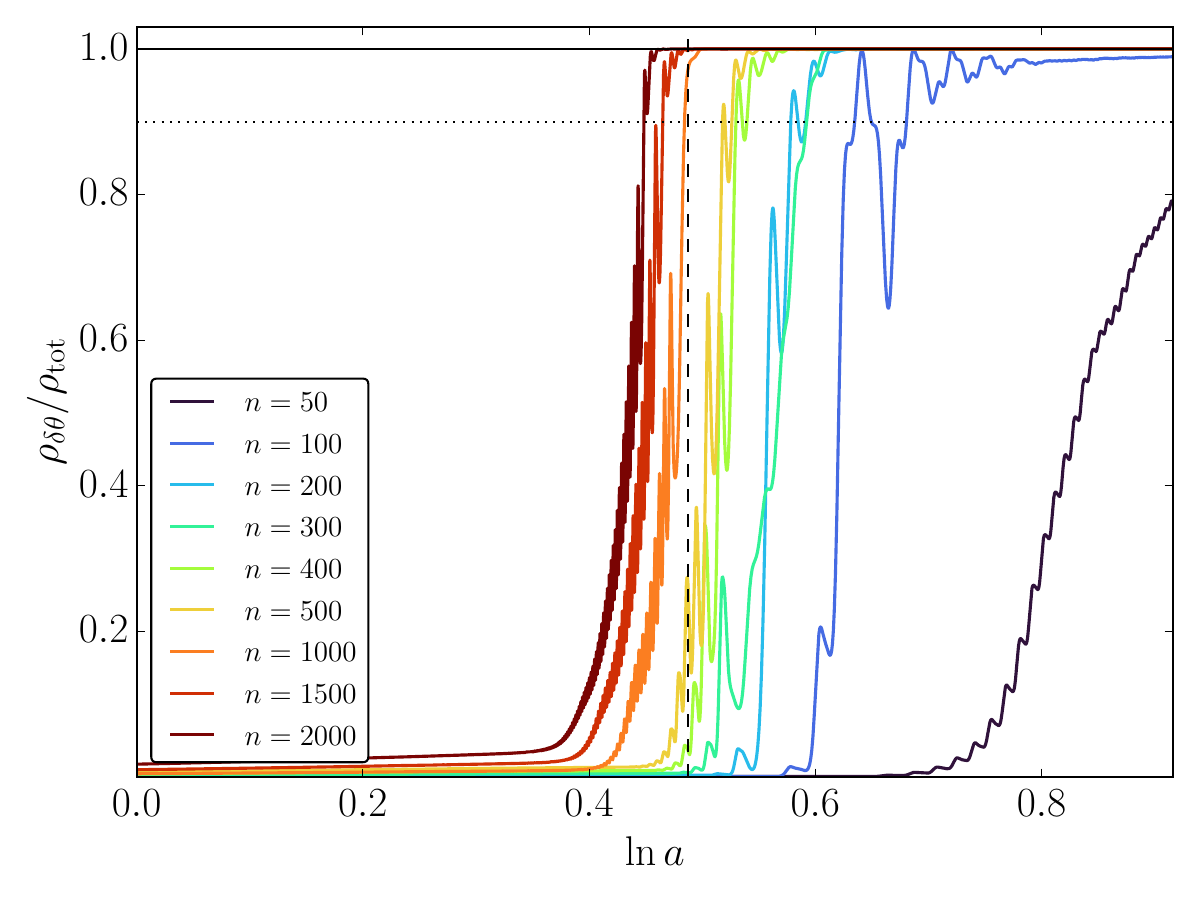}
    \caption{The evolution of the fraction of the energy in fluctuations for various values of $n$. The threshold for fragmentation defined in Eq.~(\ref{eq:frag-criterion}) is indicated in a horizontal dotted line. The vertical dashed line indicates the theoretical moment of trapping defined in Eq.~(\ref{eq:trapping-QCD}).}
    \label{fig:Fragmentation}
\end{figure}

We identify the moment of fragmentation for various values of $n$ and present them in Fig.~\ref{fig:Fragmentation}. It is apparent that as $n$ increases, fragmentation of the axion occurs earlier and earlier. This is expected as the hierarchy between the axion mass and the Hubble rate at trapping determines the efficiency of fragmentation. This notion is clear at the level of Floquet analysis presented in \cite{Eroncel:2022vjg}. Intuitively, a small $m_*$ compared to the Hubble rate delays the redshifting of the homogeneous mode so that the axion can probe the non quadratic parts of the potential for longer. We additionally observe that the value of $n$ for which fragmentation occurs at the moment of trapping is roughly $n\simeq 1000$ which is very close to the value found in previous literature~\cite{Eroncel:2022efc}, especially given how similar the results are for $n\gtrsim 1000$. The critical value of the axion decay constant in this case is $f_{\rm crit,lattice}\simeq 1.01\times 10^9\;{\rm GeV}$.

It is of course entirely possible to pinpoint both the value of $n$ as well as the critical axion decay constant corresponding to fragmentation at the moment of trapping with much greater accuracy. However, one should point out that there remains a mild sensitivity to the value of the axion initial misalignment angle at the start of the simulation. We chose to initiate the simulation at a point corresponding to a maximum of the axion potential and yet, even though the potential energy is initially a negligible fraction of the total energy, the phase choice has an impact in the late time oscillations  in Fig.~\ref{fig:Fragmentation}. As a result, the exact moment of fragmentation will always be mildly dependent on the initial phase. Our conclusion is that the lattice analysis yields results on the moment of axion fragmentation that are largely overlapping with  previous literature.
%
%
\subsection{The dilution factor}
\label{Sec:dilution-factor}
%
%

%
One of the challenging aspects of our numerical analysis is that we generally need to know the trapping temperature (\ref{eq:trapping-QCD}) in order to set our initial conditions. Such temperature is calculated analytically assuming that the late-time axion abundance matches the observed dark matter one, while disregarding the effects of fragmentation. This implies that, upon taking into account the effects of fragmentation, one may find that the true axion abundance undergoes a significant offset. In order to quantify this offset, we define the ``dilution factor" (see also Eq.~(4.24) of~\cite{Eroncel:2022efc}) as:
\begin{equation}
{\cal Z}\equiv\frac{\rho_{\rm tot,lattice}}{\rho_{\rm tot, no-frag}}\; ,
    \label{eq:dilution-factor}
\end{equation}
where $\rho_{\rm tot,lattice}$ is the total energy in the lattice analysis and  $\rho_{\rm tot, no-frag}$ is the total energy computed numerically in the absence of fragmentation. 
This dilution factor, viewed as a function of time is expected to be constant at late times as the axion perturbations redshift to the point where they become highly non-relativistic. At that stage, as pointed out in the previous section, the evolution of the perturbations is expected to match the zero-mode in the absence of fragmentation and hence the dilution factor should be very close to a constant. 

\begin{figure}[h!]
    \centering
    \includegraphics[width=0.75\linewidth]{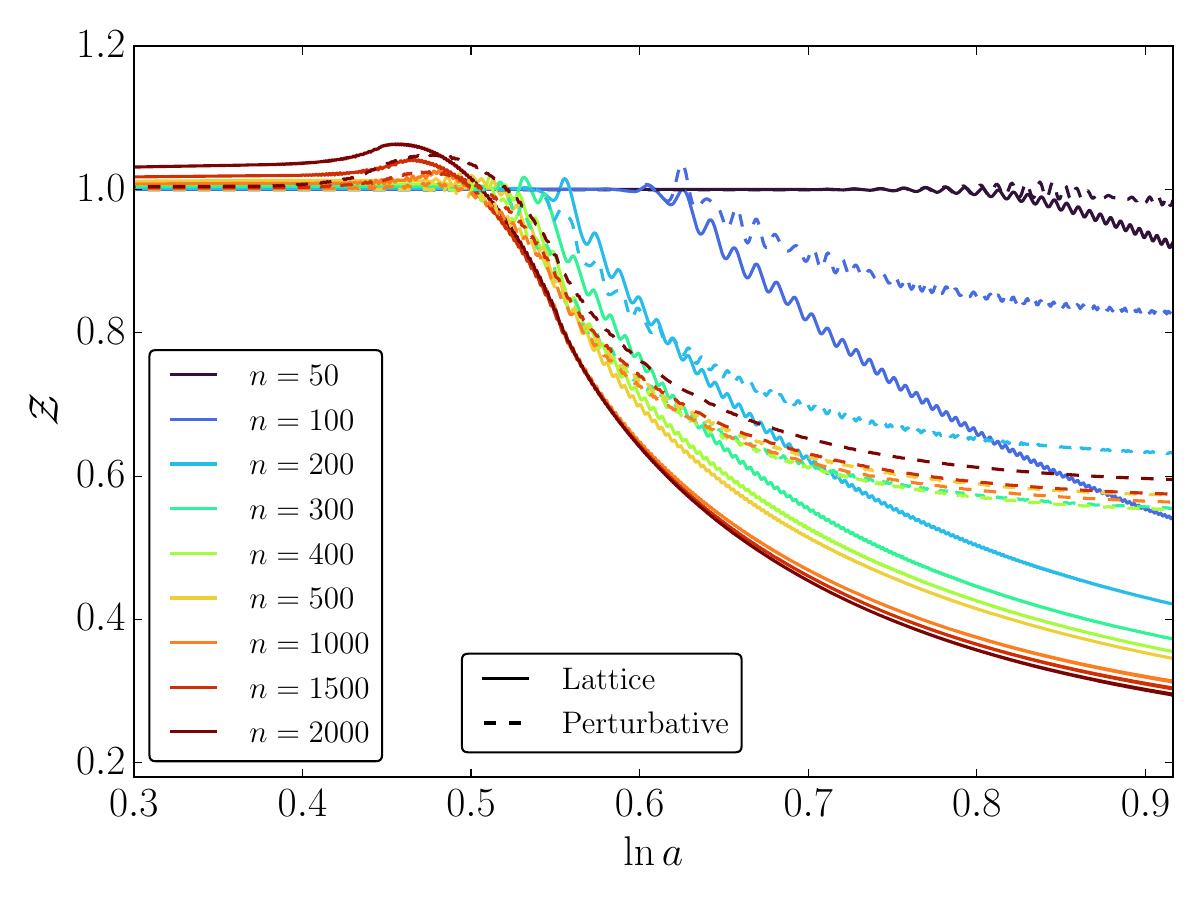}
    \caption{The evolution of the dilution factor defined in Eq.~(\ref{eq:dilution-factor}) as a function of time. We observe much smaller dilution factors on the lattice compared to the homogeneous backreaction analysis.}
    \label{fig:Dilution_comp}
\end{figure}

However, around the time of fragmentation, there can be excitement of modes which are mildly relativistic which would result in a time dependent dilution factor. We show in Fig.~\ref{fig:Dilution_comp} the evolution of the dilution factor for our set of values of $n$, from lattice simulations (solid lines) and homogeneous backreaction (dashed lines). There are several important points worth emphasizing. 

\begin{itemize}
    \item The lattice analysis yields overall a much smaller dilution factor than the homogeneous approximation. This is expected in light of the spectra we presented in the previous section. They indicated a continuous transfer of energy towards higher and higher comoving wavenumbers in the lattice result. The presence (and the scaling) of more (mildly) relativistic modes at late times implies that in the lattice case it is reasonable to expect, as compared to the zero mode and the homogeneous  analysis, an additional energy loss.
    
    \item The continuous energy transfer towards larger and larger comoving wavenumbers becomes weaker and weaker as time evolves and therefore even in the lattice analysis we observe a tendency towards the flattening of the dilution factor at late times. As a result, it is reasonable to assume that at very late times (much beyond the furthest point in time considered of our analysis) the dilution factor is constant also in the non-perturbative dynamics. 
    \item Ref.~\cite{Eroncel:2022efc} computes the dilution factor at the moment of fragmentation. As a result, it is found that for some values of the parameters the dilution factor can be greater than one. Indeed, we confirm both in the lattice as well as in the homogeneous backreaction approach that the dilution factor may be greater than one at the moment of fragmentation (indicated by the small bumps around $\ln a\simeq 0.5$ in Fig.~\ref{fig:Dilution_comp}). This is especially true  in the cases of fragmentation before trapping. However, we stress here the importance of  the post-fragmentation evolution of the dilution factor. It is the value of the dilution factor at late times that is the most informative. 
    Naturally, we cannot rule out the possibility that the dilution factor may be greater than one at late times for regions of parameters beyond the scope of our analysis.  It remains paramount to track also its post-fragmentation evolution.
\end{itemize}

To summarize, the non-perturbative lattice dynamics generally yield a dilution factor that is ${\cal O}(1)$ smaller than the homogeneous backreaction approximation and the post-fragmentation evolution of the dilution factor plays a crucial role in determining its late time value approaching a constant. In light of these results, we developed an algorithm for compensating the dilution factor and present it in the next subsection. 
%
%
\subsection{Dilution factor compensation algorithm}
\label{Sec:iteration}
%
%
\begin{figure}
    \centering
    \includegraphics[width=0.495\linewidth]{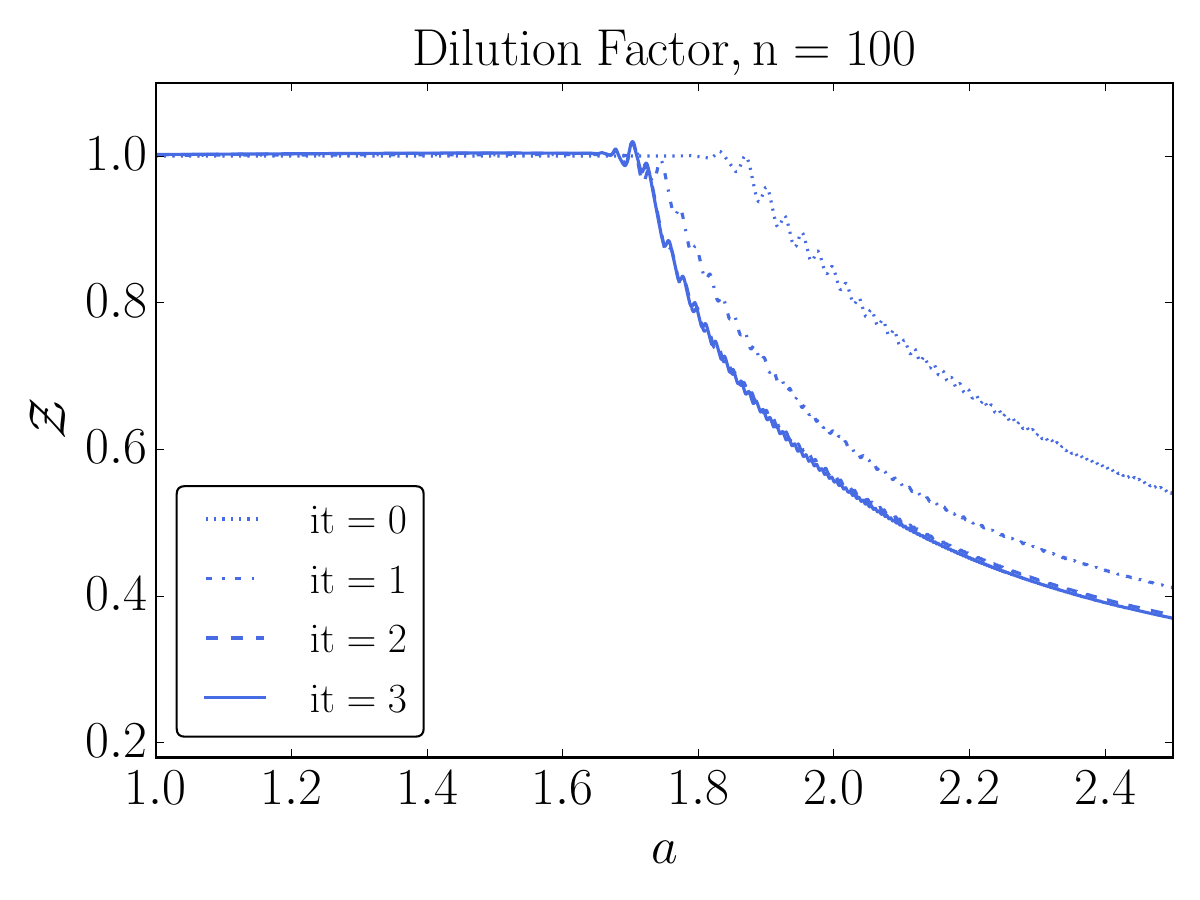}
    \includegraphics[width=0.495\linewidth]{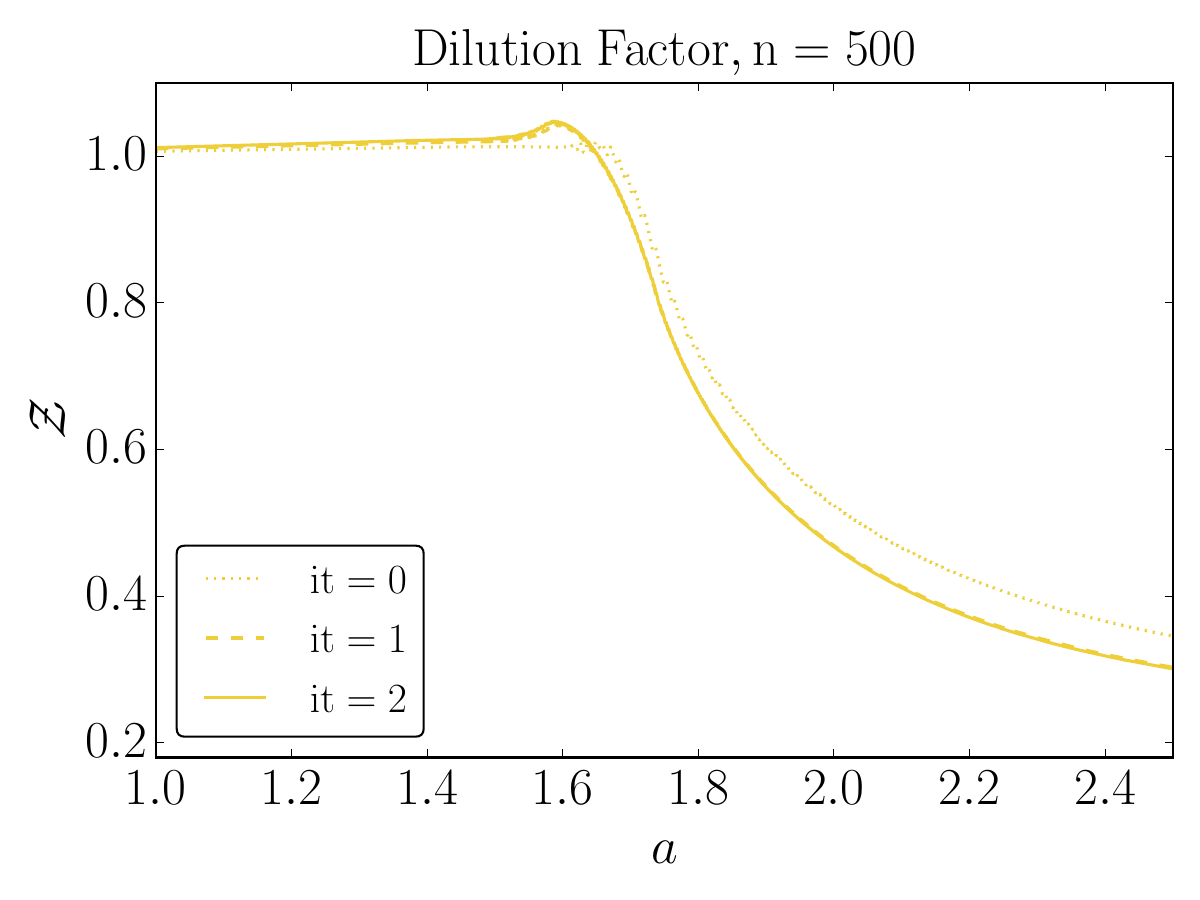}
    \caption{Evolution of the dilution factor for subsequent iterations of the compensation algorithm described in~\ref{Sec:iteration}. It is easy to see that our algorithm converges quickly to a unique late time dilution factor. The examples shown above correspond to $n=100$ (left panel) and $n=500$ (right panel) which refers to the value of $n$ only in the first iteration, which in turn fixes the mass and axion decay constant which we use in subsequent iterations.}
    \label{fig:Dilution_compensation}
\end{figure}
Given that the appeal of the scenario we have been studying relies in no small part on its ability to provide a good dark matter candidate, it is imperative to devise a way to compensate for the offset between the late-time axion abundance and the observed dark matter abundance which arises as a consequence of fragmentation. We developed an iterative algorithm which serves this function and yields, after only a few iterations, an overall axion abundance that is very close to the observed  dark matter one while taking into account the effects of fragmentation. 

At every iteration step, the algorithm entails solving the system (\ref{eq:system}) setting the initial conditions as outlined in Sec.~\ref{Sec:parametrization}. Subsequently, the dilution factor after fragmentation is fitted with a test function of the form 
\begin{equation}
    {\cal Z}_{\rm test}(a)=x \,{\rm e}^{y \,a}+z\; ,
\end{equation}
where $x$,  $y$ and $z$ are constant parameters. This fitting is necessary in order to estimate the late time constant dilution factor $\lim_{a\rightarrow \infty}{\cal Z}_{\rm test}(a)=z$. We then use this value to compensate for the trapping temperature in the next iteration using
\begin{equation}
    T_{*,i}^{\rm QCD}\simeq\left(1.18 \;{\rm GeV}\right) \left(\frac{1}{{\cal Z}_{{\rm test},\,i-1}}\right)^{-0.141}\;,
    \label{eq:trapping-QCD-mod}
\end{equation}
where ${\cal Z}_{{{\rm test},\,0}}=1$. One also must exercise caution in selecting the value of $n$ at every iteration. In the first iteration the value of $n$ is arbitrary and it fixes the mass and axion decay constant $(m_0,f)$. In order to maintain the same values in the subsequent iterations, one ought to find the corresponding $n$ value, which leaves the mass and axion decay constant invariant in light of the new trapping temperature, which can be done using Eqs.~(\ref{eq:ndef}) and~(\ref{eq:mass}).

In practice, the algorithm above compensates for the decrease in energy due to fragmentation by pushing the trapping of the axion to later and later times (lower and lower temperatures). As a result, we start off with excess energy with respect to the default scenario which is then shed due to fragmentation in such a way that the axion yields the observed dark matter abundance. In order for our algorithm to be valid, the late time dilution factor should converge to a constant value upon subsequent iterations. In Fig.~\ref{fig:Dilution_compensation} we see that indeed such a convergence is achieved after a very few iteration steps. We conclude that it should not be too difficult to compensate for the dilution factor and, at least in the case of the QCD axion, this entails  a small change to the trapping temperature with respect to the one obtained analytically. We have verified that, even after we compensate for the effect of the dilution factor, trapping dynamics occurs well before the axion potential barrier reaches its zero temperature late-time value.
%
%
\section{Conclusions}
\label{sec:conclusions}
%
%
Kinetic fragmentation has attracted considerable attention in recent years. It is viewed as a promising alternative scenario to the standard axion misalignment mechanism
for dark matter generation in the early universe. Kinetic fragmentation is not only well motivated from the perspective of UV completion \cite{Co:2019jts,Eroncel:2024rpe,Lee:2024bij}, but it also offers several advantages with respect to vacuum misalignment in terms of connections to experiments. More specifically, kinetic fragmentation can enable axions with smaller decay constants, which would otherwise underproduce dark matter, to account for the full dark matter abundance. Additionally, the rapid fragmentation of the axion enhances small-scale fluctuations, potentially giving rise to axion mini-halo formation and gravitational wave production \cite{Eroncel:2022vjg,Eroncel:2022efc}. In our scenario the prospects for detecting gravitational waves are rather low given that  the gravitational wave power generated by this mechanism depends strongly on the axion decay constant $f$ \cite{Eroncel:2022vjg}, which is constrained to be small for the QCD axion.

In light of the strong theoretical and phenomenological motivation for kinetic fragmentation, it is crucial to assess the robustness of its predictions, particularly in the presence of non-linear effects. To this end, we studied the dynamics of kinetic fragmentation using non-perturbative classical lattice simulations and compared the results with a linear approximation of the same system. For concreteness, we restricted our analysis to parameter values relevant to the QCD axion, which provides a solution to the strong CP problem of QCD. Consequently, our results are particularly important for the parameter space targeted by upcoming haloscope experiments. 

 The non-perturbative effects captured by our lattice simulation illuminate several key aspects of kinetic fragmentation. Notably, we find that the spectrum of axion fluctuations is significantly broader and more suppressed compared to the homogeneous backreaction approximation. This is expected to affect the density profile of mini-halo formation. The energy transfer we observe towards the UV implies that the axion fluctuations remain closer to the relativistic regime for longer than predicted by the homogeneous analysis. This, in turn, impacts the post-fragmentation evolution of the  total energy of the system, leading to a reduced dark matter abundance. The suppression of the axion energy at late times can be undone by increasing the initial kinetic energy of the zero-mode. We provided an algorithm to systematically adjust the initial conditions to achieve this ``compensation'' for any axion mass. Pinpointing the correct initial kinetic energy as a function of the axion mass is highly relevant from the UV completion perspective, which motivates the initial kinetic energy of the axion.

Our work can also be seen as a study of the impact of non-perturbative effects on systems exhibiting fragmentation. This approach allowed us to identify key qualitative effects missed by the homogeneous analysis. We find that the homogeneous approximation is only valid well before fragmentation; beyond this point, non-perturbative techniques become indispensable for making precise quantitative predictions about the phenomenology of this dark matter production mechanism.

%
%
\acknowledgments 
%
%
MF and AP acknowledge the “Consolidaci\'{o}n Investigadora” grant CNS2022-135590. The work of MF and AP is partially supported by the Spanish Research Agency (Agencia Estatal de Investigaci\'{o}n) through the Grant IFT Centro de Excelencia Severo Ochoa No CEX2020-001007-S, funded by MCIN/AEI/10.13039/501100011033. MF acknowledges support from the “Ram\'{o}n y Cajal” grant RYC2021-033786-I. JL and AU acknowledge the support from Eusko Jaurlaritza (IT1628-22) and the PID2021-123703NB-C21 grant funded by MCIN/AEI/10.13039/501100011033/ and by ERDF; ``A way of making Europe''. In particular AU gratefully acknowledges the support from the University of the Basque Country grant PIF20/151. The simulations of this work have been possible thanks to the computing infrastructure of the Solaris cluster at the University of the Basque Country, UPV/EHU and the Hyperion cluster from the DIPC Supercomputing Center. Part of this work was carried out during the 2025 workshop ``The Dawn of Gravitational Wave Cosmology'', supported by the Fundaci\'{o}n Ram\'{o}n Areces and hosted by the ``Centro de Ciencias de Benasque Pedro Pascual''. We thank both the CCBPP and the Fundaci\'{o}n Ram\'{o}n Areces for providing a stimulating and highly productive research environment.
%
\appendix
%
\section{Initial conditions for the perturbations}
\label{app:initial-conditions}
%
%
The initial conditions of the axion perturbations in this scenario are highly non-trivial. During radiation domination, if the axion zero-mode features a large velocity, the superhorizon perturbations of the axion are affected by the superhorizon density fluctuations. This was initially pointed out in \cite{Sikivie:2021trt,Kitajima:2021inh} and has been used explicitly in the context of kinetic fragmentation in \cite{Eroncel:2022vjg}. We defer the reader to these references for the complete derivation. Instead we will only outline the steps in this Appendix for completeness. Defining the perturbed metric as
\begin{equation}
    ds^2=a(\tau)^2\left\{-\left[1+2\Psi(\tau,\vec{x})\right]d\tau^2+\left[1+2\Phi(\tau,\vec{x})\right]\delta_{ij}dx^{i}dx^{j}\right\}\;.
\end{equation}
We assume the absence of anisotropic stress $\Psi=-\Phi$. Under these conventions, the primordial density perturbations couple to the speed of the axion zero mode and source the axion perturbations at superhorizon scales. At early times when the mass of the axion can be neglected the equation of motion of the perturbations takes the form
\begin{equation}
    \theta_k''+2\frac{a(\tau)'}{a(\tau)}\theta_k'+k^2\theta_k=-4\Phi_k'\Theta'
\end{equation}
It is useful to rewrite this equation in terms of variable $x\equiv k\tau/\sqrt{3}$
\begin{equation}
    \frac{d^2\theta_k}{dx^2}+\frac{2}{x}\frac{d\theta_k}{dx}+3\theta_k=-4\frac{d\Phi_k}{dx}\frac{d\Theta}{dx}\equiv{\cal F}(x)
\end{equation}
This is an inhomogeneous differential equation whose solution can be written with the Green function method. First the homogeneous solution takes the form
\begin{equation}
    \theta^{\rm hom}_k(x)=\frac{1}{x}\left[c_1\cos(\sqrt{3}x)+c_2\sin(\sqrt{3}x)\right]\;,
\end{equation}
while the particular solution associated to the inhomogeneous term is given by
\begin{equation}
    \theta^{\rm par}_k=\theta_s(x)\int^x dx' \frac{\theta_c(x'){\cal F}(x')}{W\left[\theta_c(x'),\theta_s(x')\right]}-\theta_c(x)\int^x dx'\frac{\theta_s(x'){\cal F}(x')}{W\left[\theta_c(x'),\theta_s(x')\right]}
\end{equation}
where $\theta_c=\cos(\sqrt{3}x)/x$ and $\theta_s=\sin(\sqrt{3}x)/x$. The Wronskian of the homogeneous solutions is $W=\sqrt{3}/x^2$. 

We set initial conditions at an early time $x_{\rm early}$ at which all perturbations were superhorizon. Note that this time should not be confused with the starting time of our simulation corresponding to $T_{\rm ini}$. The inhomogeneous term can then be written as\footnote{Here we used the solution of the Newtonian potential in a radiation dominated universe $\Phi_k(x)=3\Phi(0)\left[\frac{\sin(x)-x\cos(x)}{x^3}\right]$. Where $\Phi_k(0)=\frac{3}{2}{\cal R}_k(0)$ and the comoving curvature perturbation is given by inflation as $\langle|{\cal R}_k(0)|^2\rangle=\frac{2\pi^2}{k^3}A_s$. Our value for $A_s=2.1\times 10^{-9}$ is in agreement with \cite{Planck:2018vyg}.}
\begin{equation}
    {\cal F}(x)=-12\Phi_k(0)\frac{\dot{\Theta}_{\rm early}}{H_{\rm early}}\frac{\eta_{k,{\rm early}}}{\eta_{k}^2}\left[\frac{\sin(x)}{x^2}+\frac{3\cos(x)}{x^3}-\frac{3\sin(x)}{x^4}\right]\;.
\end{equation}
This allows us to write the full solution as 
\begin{equation}
    \theta_k(x)=\frac{1}{x}\left\{\cos\left(\sqrt{3}x\right)\left[c_1-{\cal I}_s(x)+{\cal I}_s(x_{\rm early})\right]+\sin\left(\sqrt{3}x\right)\left[c_2+{\cal I}_c(x)-{\cal I}_c(x_{\rm early})\right]\right\}
\end{equation}
where the integrals ${\cal I}_{c,s}$ are defined
\begin{equation}
    {\cal I}_{c,s}(x)=\int^x dx'\frac{\theta_{c,s}(x'){\cal F}(x')}{W\left[\theta_c(x'),\theta_s(x')\right]}\;.
\end{equation}

At this stage Ref.~\cite{Eroncel:2022vjg} imposes that at early times the perturbations are adiabatic. This fixes the constants $c_1$ and $c_2$ to be
\begin{equation}
    c_1=-\frac{1}{2}x_{\rm early}\frac{\Theta'_{\rm early}}{a_{\rm early}H_{\rm early}}\Phi_k(0)\;\;\;,\;\;\;c_2=0\;.
\end{equation}
Using the early time scaling $\Theta'_{\rm early}\sim a^{-2}$ and $a H\sim a^{-1}$ we find the approximately early time solution at superhorizon scales to be
\begin{equation}
    \theta_{k,{\rm ad}} \simeq-\frac{1}{2}\frac{\Theta'}{a\, H}\Phi_k(0)\cos\left(k \tau\right)\;,
    \label{eq:initial-conditions}
\end{equation}
Where $\Theta'$ and $a \, H$ can be evaluated at any moment after $x_{\rm early}$. Since we are evaluating the code initial conditions at a moment when all the modes are well within the horizon, one should in principle derive a more general solution that considers the effects of the rest of the terms beyond the homogeneous one which become important at subhorizon scales. Such contributions however only affect the initial conditions up to ${\cal{O}}(1)$ corrections so in practice keeping only Eq.~(\ref{eq:initial-conditions}) yields a sufficient answer up to the sensitivity required in our simulation.

Ref.~\cite{Eroncel:2022vjg} makes a further approximation by disregarding the $\cos(k\tau)$ term in Eq.~(\ref{eq:initial-conditions}) since that is a rapid oscillation with amplitude of one deep inside the horizon. In practice we evaluate the initial conditions using Eq.~(\ref{eq:initial-conditions}), disregarding the cosine, at moment $T_{\rm ini}$ when we initialize our numerical simulations.

Our choice of adiabatic fluctuations effectively make the inhomogeneous contribution to the full equation of motion irrelevant. This, at face value appears odd, since the only way the axion perturbation can be influenced by the primordial curvature perturbation is through the inhomogeneous term. As a result we also investigated the choice $c_1=c_2=0$ which effectively switches off the homogeneous terms leaving the inhomogeneous solution to control the early time solutions of the axion perturbation. The resulting expressions are too complicated to recreate here, but the final result is of the same order of magnitude as in the adiabatic choice outlined before. We have verified that this choice of initial conditions yields nearly exactly the same numerical results as the ones corresponding to the adiabatic initial conditions.

%
%
\section{Numerically simulating the homogeneous backreaction case}
\label{app:homogeneous-backreaction}
%
%
\begin{figure}
    \centering
    \includegraphics[width=0.495\linewidth]{EnDensities_homo_n100.pdf}
    \includegraphics[width=0.495\linewidth]{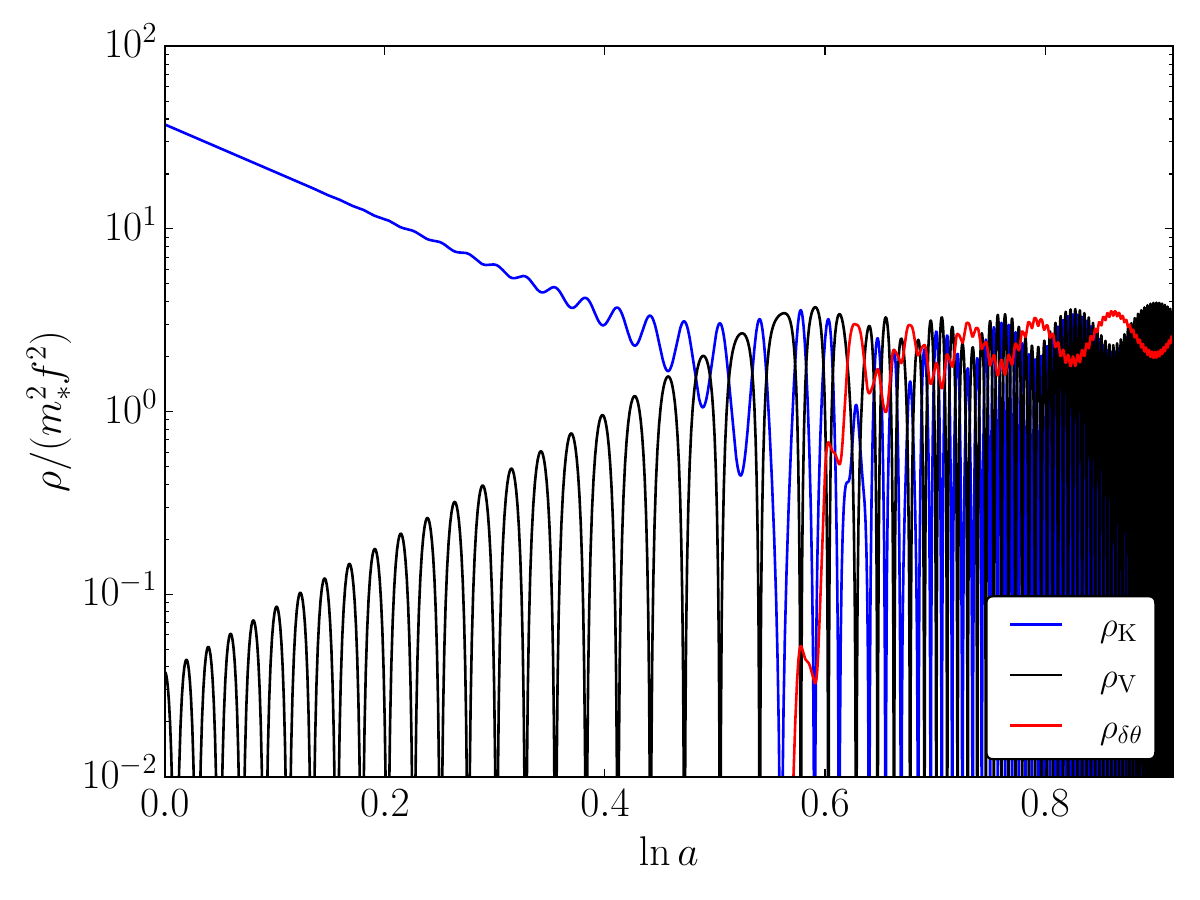}
    \caption{\textit{Left panel}: Evolution of the kinetic and potential energy densities of the zero mode disregarding the effects of perturbations. \textit{Right panel}: Evolution of the system of eq.~(\ref{eq:eomnume})   Sec.~\ref{Sec:homogeneous-backreaction}.}
    \label{fig:hom-bck-naive}
\end{figure}
A naive numerical computation of the system in Eq.~(\ref{eq:eomnume}) up to as late times as possible typically either yields two possible results. The system either exhibits oscillations resembling the example of Fig.~\ref{fig:hom-bck-naive} or displays a spurious divergence of the energy in fluctuations depending on the choice of parameters. In the either case, such a result cannot be physical and a careful examination of the equations of motion reveals the reason for this spurious result. Eq.~(\ref{eq:eomnume}) can be rearranged as follows
\begin{equation}
    \ddot{\Theta}+3H\dot{\Theta}+m(T)^2 \sin(\Theta)\left[1-\alpha(t)\right]=0
    \label{eq:naive}
\end{equation}
where
\begin{equation}
    \alpha(t)\equiv\frac{1}{2}\int\frac{d^3k}{(2\pi)^3}|\theta_k|^2
\end{equation}
This rewriting of the equation reveals that the backreaction term acts as an effective potential which initially, when $\alpha(t)<1$, effectively flattens the potential slope. On the other hand, when the perturbations grow a lot, $\alpha(t)$ can be come ${\cal O}(1)$ and in that case the potential flips sign, turning the maxima of the potential into minima and vice versa. This forces the axion to oscillate around the potential maximum which is precisely where the perturbations are produced most rapidly. This creates a runaway effect which is a consequence of the breakdown of perturbation theory

The fact that the backreaction term can be absorbed in the manner shown in Eq.~(\ref{eq:naive}) is of course an accidental side effect of the sinusoidal potential of the axion. Additionally, such an absorption would not be possible if we kept higher order terms in our backreaction analysis, since the next order term is proportional to $V^{(4)}$ which yields a cosine as opposed to a sine. 

The flipping of the sign of the potential at around $\alpha(t)=1$ signals the need to consider higher order terms in our backreaction and therefore that moment can be conservatively be taken to be the moment our system moves beyond perturbative control. Unfortunately, by that time, the energy in fluctuations is not high enough to declare fragmentation using the strict definition we use in the lattice case (see Eq.~(\ref{eq:frag-criterion})). As a result, in the homogeneous backreaction approximation, we simply declare fragmentation at the moment when perturbativity breaks down. 

After fragmentation, we continue to solve the system of fluctuations beyond, assuming the zero-mode to be stuck in its minimum. This is precisely the expected late time evolution of the system since sooner or later the residual zero-mode is expected to oscillate around the minimum with very small amplitude. This strategy allows us to stitch together early time and late time perturbative solutions missing only the non-perturbative dynamics in the short period between them.

\bibliography{arxiv_1,arxiv_1_manual}
\end{document}